%% file: main.tex
\documentclass[format=acmsmall, screen=false, review=false]{acmart}

\AtBeginDocument{%
  \providecommand\BibTeX{{%
    \normalfont B\kern-0.5em{\scshape i\kern-0.25em b}\kern-0.8em\TeX}}}

\usepackage{fontawesome}
\usepackage{xspace}
\usepackage{siunitx}
\usepackage{enumitem}
\usepackage{balance}
\usepackage{graphics}
\usepackage{color, colortbl, soul, xcolor}
\usepackage{subcaption}
\usepackage{fixltx2e}
\usepackage{mathtools}
\usepackage{makecell}
\usepackage{amsmath}
\usepackage[title]{appendix}
\usepackage{bm}
\usepackage{algorithm}
\usepackage{algorithmic}

\newcommand\algorithmicprocedure{\textbf{function}}
\newcommand{\algorithmicendprocedure}{\algorithmicend\ \algorithmicprocedure}
\makeatletter
\newcommand\PROCEDURE[3][default]{%
  \ALC@it
  \algorithmicprocedure\ \textsc{#2}(#3)%
  \ALC@com{#1}%
  \begin{ALC@prc}%
}
\newcommand\ENDPROCEDURE{%
  \end{ALC@prc}%
  \ifthenelse{\boolean{ALC@noend}}{}{%
    \ALC@it\algorithmicendprocedure
  }%
}
\newenvironment{ALC@prc}{\begin{ALC@g}}{\end{ALC@g}}
\makeatother

\def\S{Sec.\xspace}

\def\ie{\textit{i.e.,}\xspace}
\def\etal{\textit{et~al.}\xspace}

\def\eg{\textit{e.g.,}\xspace}
\def\incl{\textit{incl.}\xspace}

\def\vs{\textit{vs.}\xspace}
\def\cf{\textit{cf.}\xspace}

\newcommand{\blue}[1]{\textcolor{black}{#1}}

\setcopyright{rightsretained} 
\acmJournal{TOCHI}
\acmYear{2024} 
\acmVolume{31} 
\acmNumber{5} 
\acmArticle{1} 
\acmMonth{10}
\acmDOI{10.1145/3694681}

\author{Chen Chen}
\authornote{This work was conducted while Chen was interning at Adobe Research in 2023.}
\orcid{0000-0001-7179-0861}
\affiliation{%
  \department{Computer Science and Engineering}
  \institution{University of California San Diego}
  \city{La Jolla}
  \state{CA}
  \country{United States}
}
\email{chenchen@ucsd.edu}

\author{Cuong Nguyen}
\orcid{0000-0001-9234-9960}
\affiliation{%
  \institution{Adobe Research}
  \city{San Francisco}
  \state{CA}
  \country{United States}
}
\email{cunguyen@adobe.com}

\author{Thibault Groueix}
\orcid{0000-0002-7984-8252}
\affiliation{%
  \institution{Adobe Research}
  \city{San Francisco}
  \state{CA}
  \country{United States}
}
\email{groueix@adobe.com}

\author{Vladimir G. Kim}
\orcid{0000-0002-3996-6588}
\affiliation{%
  \institution{Adobe Research}
  \city{Seattle}
  \state{WA}
  \country{United States}
}
\email{vokim@adobe.com}

\author{Nadir Weibel}
\orcid{0000-0002-3457-4227}
\affiliation{%
  \department{Computer Science and Engineering}
  \institution{University of California San Diego}
  \city{La Jolla}
  \state{CA}
  \country{United States}
}
\email{weibel@ucsd.edu}

\ccsdesc[500]{Human-centered computing~Visualization systems and tools}

\keywords{3D Design Feedback, Reference Images, Tools for Asynchronous Design Collaborations, Applications of Generative AI and Vision-Language Foundation Models}

\begin{document}

\def\sysname{MemoVis}

\def\formativeStudyRecruitingSource{Nissan Design America}


\title{\sysname: A GenAI-Powered Tool for Creating Companion Reference Images for 3D Design Feedback}

\begin{teaserfigure}
    \includegraphics[width=\textwidth]{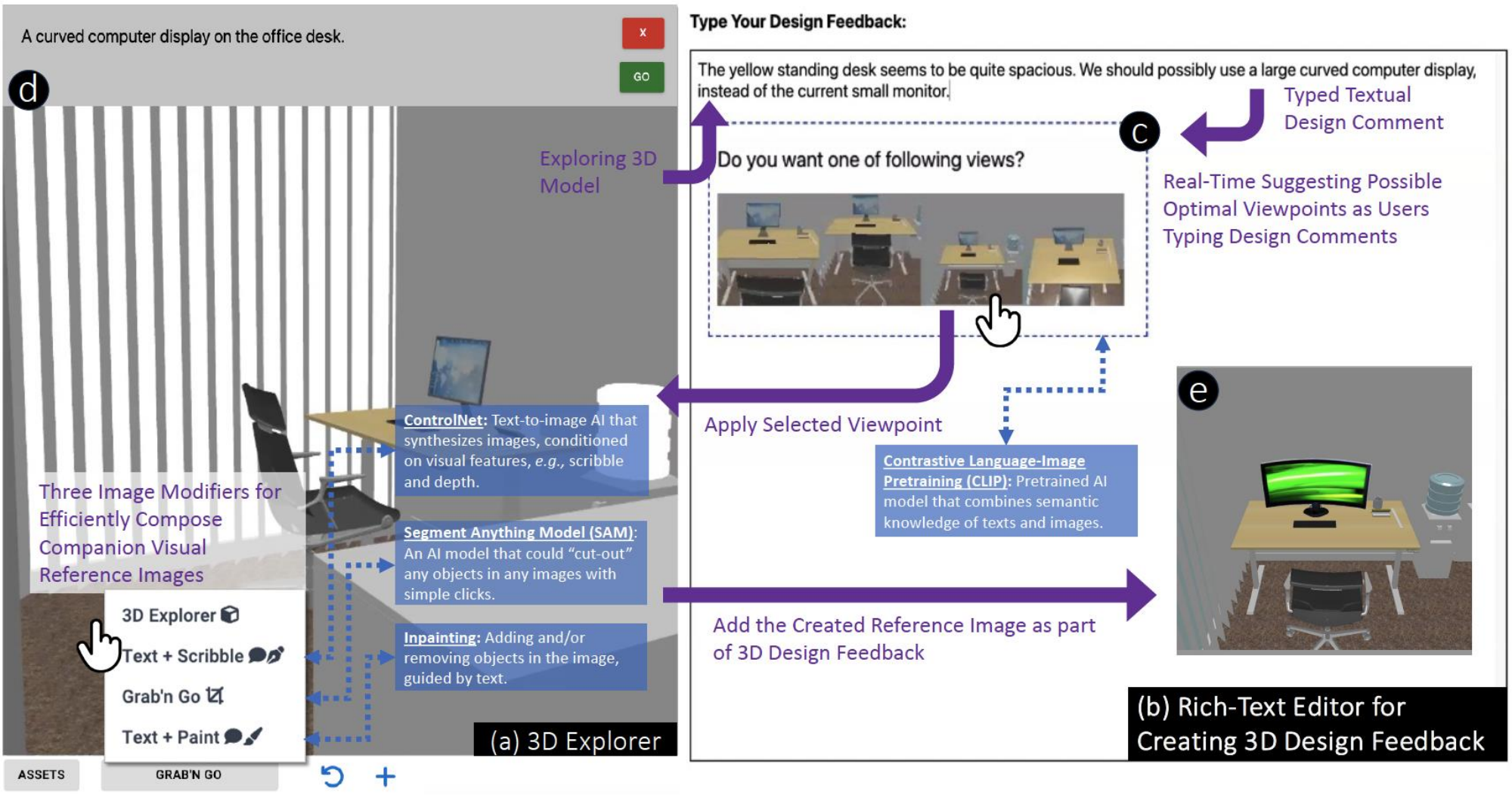}
    \caption{\sysname~is a browser-based text editor interface that assists feedback providers to create companion reference images for 3D design. Feedback providers can (a) explore the 3D model in a 3D content viewer and (b) type the textual feedback comments in a rich-text editor interface; (c) a real-time viewpoint suggestion anchors the textual design comments with a camera viewpoint; (d) a textual prompt could be used to guide the AI generated images. Three types of images modifiers enable feedback providers to efficiently compose companion visual reference images; (e) the visualized reference image reflects the gist of the textual design comments and can be used as part of memo for professional designers to instantiate the feedback.}
    \label{fig::teaser}
\end{teaserfigure}

\input{0-abstract}

\pagenumbering{arabic}
\settopmatter{printfolios=true}

\maketitle


\input{1-introduction}

\input{2-related}

\input{3-formative-study}

\input{4-system}

\input{5-evaluation}

\input{6-discussions}

\input{7-conclusion}
\input{ack}


\bibliographystyle{ACM-Reference-Format}
\bibliography{reference}

\clearpage
\begin{appendices}

\input{A01-ethical-disclaimer}
\input{A03-controlnet-depth-scribble}

\clearpage
\input{A02-algorithm}

\input{A04-pretrained-models}
\input{A05-participants}
\input{A06-study-task}

\clearpage
\input{A07-modifier-usages}

\clearpage
\input{A08-codebook-online-analysis}
\end{appendices}
\clearpage


\end{document}

%% file: 0-abstract.tex
\begin{abstract}
Providing asynchronous feedback is a critical step in the 3D design workflow.
A common approach to providing feedback is to pair textual comments with companion reference images, which helps illustrate the gist of text. 
Ideally, feedback providers should possess 3D and image editing skills to create reference images that can effectively describe what they have in mind.
However, they often lack such skills, so they have to resort to sketches or online images which might not match well with the current 3D design.
To address this, we introduce \emph{\sysname}, a text editor interface that assists feedback providers in creating reference images with generative AI driven by the feedback comments. 
First, a novel real-time viewpoint suggestion feature, based on a vision-language foundation model, helps feedback providers anchor a comment with a camera viewpoint. 
Second, given a camera viewpoint, we introduce three types of image modifiers, based on pre-trained 2D generative models, to turn a text comment into an updated version of the 3D scene from that viewpoint. 
We conducted a within-subjects study with $14$ feedback providers, demonstrating the effectiveness of \sysname. 
The quality and explicitness of the companion images were evaluated by another eight participants with prior 3D design experience.

\end{abstract}

%% file: 1-introduction.tex
\section{Introduction}\label{sec::intro}
Providing asynchronous feedback is a critical step in the 3D design workflow. 
Exchanging feedback allows all stakeholders such as collaborators and clients to review the design collaboratively, highlight issues, and propose improvements~\cite{Gibbons2016, CC3DModel2023, AsyncFeedback2017}.
However, creating effective and actionable feedback is often challenging for 3D design. 
Feedback providers typically convey suggestions about the 3D design via texts.
This practice is similar to doing design review in many 2D domains such as videos~\cite{pavel2016vidcrit,nguyen2017collavr} and documents~\cite{warner2023slidespecs}. 
However, browsing \blue{3D models} requires users to navigate information using a viewing camera with six \textbf{D}egrees \textbf{o}f \textbf{F}reedom~($6$DoF). 
Typically, it is more tedious for users to convey where and how changes should be applied on a 3D canvas compared to writing feedback on 2D media.
Additionally, describing certain types of design changes, such as material and texture, can be challenging without extensive 3D knowledge. 
These issues make it especially challenging for individuals with different skill levels to collaborate effectively, like when a designer and a client need to exchange feedback.

To make feedback more instructive, attaching reference images to the textual feedback comments is a common approach for feedback providers to illustrate the texts and externalize the critiques~\cite{Gibbons2016, Barnawal2017}.
The process of creating reference images could also inspire feedback providers to find alternative design problems and generate more insights~\cite{Kang2018, Holinaty2021}.
Ideally, feedback providers should possess basic 3D and image editing skills to create reference images that can effectively describe their thoughts. 
But 3D design is time-consuming, and some users might not have the proficiency to express their thoughts using a 3D software~\cite{CC3DModel2023}. 
As a result, they often resort to sketches or images found on the internet.
For example, when designing the appearance of a 3D bedroom, it might be quicker for a client to search for bedroom images on websites such as Pinterest than to use 3D software like Blender to adjust model's geometry, materials, and textures. 
However, searching for reference images on the internet can also be challenging. 
Finding images that precisely match the viewpoint and 3D structure of the current 3D model is time consuming and often nearly impossible.
The disparity can lead to misunderstandings as designers try to understand the feedback. 
Moreover, online search engines often yield images in similar styles due to various biases in indexing algorithms, potentially causing bias in the feedback and influencing it in unintended ways~\cite{Otterbacher2018}.

Recent \textbf{Gen}erative \textbf{A}rtificial \textbf{I}ntelligence (GenAI) and \textbf{V}ision-\textbf{L}anguage \textbf{F}oundation \textbf{M}odels (VLFMs) offer unique opportunities to address these challenges. 
Text-to-image generation workflow enabled by generative models have been deployed in commercial tools (\eg~Firefly~\cite{AdobeFirefly} and Photoshop~\cite{Photoshop, photoshopAI}) and are being actively studied in HCI with applications in both 3D ~\cite{Liu2023} and 2D ~\cite{Cai2023, Son2023GenQuery} design ideation. 
However, it remains unclear how these text-to-image GenAI models may support the reference image creations in the 3D design \textit{review} workflow.
Using editing tools like Photoshop~\cite{Photoshop, photoshopAI} can create high-fidelity reference images. 
However, this process is time-consuming and requires feedback providers to possess professional image editing skills.
While it is also feasible for feedback providers to generate reference images using existing text-to-image GenAI tools (\eg~\cite{AdobeFirefly, Liu2023, BlenderCopilot}), the synthesized images are usually not contextualized on the current design, making it hard for the designers to interpret the feedback.
Our formative studies indicates that additional controls for 3D scene navigation and alignment of the generated output with the 3D scene structure are crucial for reviewers to create effective reference images.

To explore how text-to-image GenAI can be integrated into the 3D design review workflow, we introduce \emph{\sysname}, a novel browser-based text editor interface for feedback providers to easily create reference images for textual comments. 
Fig.~\ref{fig::teaser} presents an overview of the workflow. 
\sysname~ enables novice users to write textual feedback, quickly identify relevant camera views of the 3D scene, and use text prompts to construct reference images. 
Importantly, these images are aligned with the chosen 3D view, enabling feedback providers to more effectively illustrate their intentions in their written comments.

\sysname~ realizes this by introducing a real-time viewpoint suggestion feature to help feedback providers anchor a textual comment with possible associated camera viewpoints.
\sysname~ also enables users to generate images using text prompts, based on the depth map of the chosen camera viewpoint. 
To provide users with more controls over the generation process, \sysname~ offers three distinct modifier tools that complement text prompt input. 
The \textit{text + scribble modifier} allows users to focus the generation on a specific object in the 3D scene. 
The \textit{grab’n go modifier} assists in composing objects from the generated images into the current 3D view. 
Lastly, the \textit{text + paint modifier} utilizes inpainting~\cite{Rombach2022SD} to aid users in making minor adjustments and fine-tuning the generated output.

To understand the design considerations of \sysname, we conducted two formative studies by interviewing two professional designers and analyzing real-world online 3D design feedback. 
With three key considerations identified from the formative studies, we then prototyped \sysname~ by leveraging recent pre-trained GenAI models~\cite{Zhang2023ControlNet, Zhao2023uni, Kirillov2023SAM, Wang2023InstructEdit}. 
Through a within-subjects study with $14$~participants, we demonstrated the feasibility and effectiveness of using \sysname ~ to support an easy workflow for visualizing 3D design feedback.
A second survey study with eight participants with prior 3D design experience demonstrated the straightforwardness and explicitness of using reference images created by \sysname ~to convey 3D design feedback.

In summary, our research around \sysname~ explores a potential path and solution to integrate text-to-image GenAI into the 3D design review workflow.
Our key contributions include:

\vspace{4px}
\begin{itemize}[noitemsep, topsep=0pt, leftmargin=*]

\item {\bf Formative studies}, exploring \textbf{(i)} the integration of GenAI into the companion image creation process for 3D design feedback and \textbf{(ii)} the characteristics of real-world 3D design feedback.

\item {\bf Design of \sysname}, a browser-based text editor interface with a viewpoint suggestion feature and three image modifiers to assist feedback providers to create and visualize 3D design feedback.

\item {\bf User studies}, analyzing the user experience with \sysname, as well as the usefulness and explicitness of the reference images created by \sysname ~to convey 3D design feedback.

\end{itemize}

%% file: 2-related.tex
\section{Related Work}\label{sec::related}

This section discusses the key related work while designing \sysname.
We first look into prior research that explored the supporting tools for creating design feedback (Sec.~\ref{sec::related::feedback}). 
We then introduce multiple GenAI and VLFMs related to \sysname ~(Sec.~\ref{sec::related::genai}), and discuss how they could be integrated into 2D and 3D design workflow (Sec.~\ref{sec::related::integrate}).

\subsection{Creating Effective Design Feedback}\label{sec::related::feedback}

Designers often receive feedback along the progress of their design from their collaborators, managers, clients, or even online community~\cite{Krause2017}.
Feedback allows designers to gather external perspectives to avoid mistakes, improve the design and examine whether the design meets the objectives~\cite{Gibbons2016}.
Feedback can also help designers foster new insights and creativity~\cite{Nijstad2006}.

In nearly all creative design processes, asynchronous feedback serves as a critical medium to communicate ideas between designers and various stakeholders~\cite{AsyncFeedback2017}.
The modalities of feedback could affect the communication efficiencies~\cite{Linsey2011}, and the feedback interpretability~\cite{Easterday2007}.
While creating text-only feedback is an easy and widely-adopted method, using visual references is often more effective.
Prior works have explored the integration of reference images with textual feedback.
Herring~\etal~\cite{Herring2009} demonstrated the importance of using reference images during client-designer communications, creating a more effective way to enable designers to internalize client needs.
Paragon~\cite{Kang2018} argued that the visual examples could encourage feedback providers to create more specific, actionable, and novel critique.
This finding guides an interface design that allows feedback providers to browse examples for visual poster design using metadata.
Robb~\etal~\cite{Robb2015} showed a visual summarization system that could crowd-source a small set of representative image as feedback, which could then be consumed at a glance by designers.

Similar to 2D visual design, the system for creating feedback is also urgently needed for \emph{3D design}.
In sectors like the manufacturing industry, the capability to provide feedback and comments is becoming an essential feature in today's 3D \textbf{D}esign \textbf{F}or \textbf{M}anufacturability (DFM) tools.
Many existing research and 3D software have designed features for feedback providers to create textual comments and draw annotations in a specific viewpoint, or on the 3D model directly.
The ModelCraft demonstrates how  freehand annotations and edits can help in the ideation phase during early 3D design stage~\cite{Song2009}.
Professional DFM software, \eg~ Autodesk Viewer, allows adding textual comments of a specific viewpoint and markups to specific parts of 3D assets~\cite{AutodeskViewerFeedback}.
Browser-based tools such as TinkerCAD~\cite{TinkerCAD} also enable feedback providers with less professional 3D skills to add textual comments.
As \textbf{V}irtual \textbf{R}eality~(VR) headsets advance, recent research has also delved into feedback creations within the context of VR-based 3D design review workflow~\cite{Wolfartsberger2019}.

While textual feedback is simple to create, and might be useful in many cases, Bernawal~\etal~\cite{Barnawal2017} showed that the graphical feedback could significantly improve performance and reduce mental workload for design engineers compared to textual and no feedback in manufacturing industry settings. 
However, creating reference images is tedious and challenging.
For certain design with less-common perspective, searching reference images that are well matched with the specific viewpoints is time consuming and sometimes nearly impossible. 
While rapid sketching or using image editing tools might work, such process is tedious and needs feedback providers to have professional image editing and 3D skills --- an impractical expectation for many stakeholders such as managers and clients.
\sysname ~ shows a novel approach that aims to leverage the power of recent GenAI to help feedback providers easily create companion reference images for 3D design feedback.

\subsection{Generative AI (GenAI) and Vision-Language Foundation Models (VLFMs)}\label{sec::related::genai}
Language and vision serve as two primary channels for information~\cite{Wu2023}.
Recent AI research has advanced the capabilities of visual and text-based foundation models, many of which have been successfully integrated into commercially available products.
Since the introduction of Generative Adversarial Network~\cite{Goodfellow2014} and Deep Dream~\cite{DeepDream2015, Szegedy2014}, many recent GenAI models, such as Stable Diffusion~\cite{Rombach2022SD}, Midjourney~\cite{Midjourney} and DALL$\cdot$E $3$~\cite{dalle3}, can understand textual prompts and generate images, using models pre-trained by billions of text-image pairs.
Beyond text-to-image generations, several VLFMs attempt to bring natural language processing innovations into the field of computer vision.
The \textbf{C}ontrastive \textbf{L}anguage-\textbf{I}mage \textbf{P}re-training (CLIP) model demonstrates a ``zero-shot'' capabilities to use texts to achieve various image classification tasks, without the need for directly optimizing the model for a specific benchmark~\cite{CLIP, Radford2021}.
The CLIP model also possesses capabilities to represent text and image embeddings in the same space, allowing for direct comparisons between the two modalities~\cite{CLIP}.
The \textbf{B}ootstrapping \textbf{L}anguage-\textbf{I}mage \textbf{P}re-training (BLIP)~\cite{Li2022, Li2023} model is another example of pre-trained VLFMs that can perform a wide variety of multi-modal tasks, such as visual question answering and image captioning.
Grounding DINO~\cite{Liu2023} shows how the transformer-based detector DINO can be integrated into grounded pre-training, to detect arbitrary objects with human input. 
Using Grounding DINO~\cite{Liu2023} and the \textbf{S}egment \textbf{A}nything \textbf{M}odel (SAM)~\cite{Kirillov2023SAM} unlocks the opportunities for inferring segmentation mask(s) based on text input.
Similarly, other pre-trained models, \eg~ Tag2Text~\cite{Huang2023} and RAM~\cite{Zhang2023}, show to generate textual tags with input images.

While text-guided image synthesis is promising, generating images just based on texts may fail to satisfy users' needs due to lack of additional ``control''.
ControlNet~\cite{Zhang2023ControlNet} shows the feasibility of adding additional control to text-to-image diffusion models. 
They show how large diffusion models such as Stable Diffusion can be augmented by additional conditional input such as edges and depths.
This opens possibilities for many follow-up works, \eg~Uni-ControlNet~\cite{Zhao2023uni} and LooseControl~\cite{Bhat2023LooseControl}, that demonstrate the integration of multimodal conditions.
InstructEdit~\cite{Wang2023InstructEdit} and early work like EditGAN~\cite{Liang2021EditGANA} show how users can perform fine-grained editing based on text-instruction.

While innovating GenAI and VLFMs models is \emph{out} of our scope, \sysname ~ contributes a novel interaction workflow to enable feedback providers easily creating reference images for 3D design feedback by leveraging the strengths of today's GenAI models.

\subsection{GenAI-Powered 2D and 3D Design Workflow}\label{sec::related::integrate}
Previous research has explored novel techniques for discovering the design space and controlling the generation process of image through \textbf{G}enerative \textbf{A}dversarial \textbf{N}etwork~(GAN)-based GenAI~\cite{Goodfellow2014}, resulting in an increased efficiency of various visual design workflow.
%
For example, Zhang~\etal~\cite{Zhang2021} illustrates how a selection of image galleries may be generated using a novel sampling methods, along with an interactive GAN exploration interface.
Koyama~\etal~\cite{Koyama2017, Koyama2020} proposes the Bayesian optimization-based approach that allows designers to search and discover the design space through a set of slider bars.
Additionally, much prior research proposes novel interaction experience that allow users to input additional control.
%
For example, GANzilla~\cite{Evirgen2022} further demonstrates how iterative scatter/gather techniques~\cite{Pirolli1996} allow users to discover editing directions --- the user-defined control to steer generative models to create content with different characteristics. 
Follow-up research, GANravel~\cite{Evirgen2023}, shows the techniques of global editing (by adding weights to example images) and local editing (with the scribbled masks) for disentangling editing directions (\ie~ achieve user-defined control while ensuring that unintended attributes remain unchanged in the target image).
GANCollage~\cite{Wan2023} shows a StyleGAN-driven~\cite{Karras2019, Karras2020} mood board that allows users to define possible controls through sticky notes.

Recent advances in text-to-image GenAI have significantly benefited previous research on integrating these technologies into 2D visual design workflows.
For example, the feature of ``Generative fill''~\cite{GenFill2023} introduced by Photoshop shows how texts with simple scribbling could easily in-paint and out-paint the target image (\eg~adding and/or removing objects).
Reframer ~\cite{Lawton2023} demonstrates a novel human-AI co-drawing interface, where the creator can use a textual prompt to assist with the sketching workflow.
The study conducted by Ko~\etal~\cite{Ko2023} demonstrates versatile roles of text-to-image GenAI for visual artists from $35$ distinct art domains to support automating art creation process, expanding ideas, and facilitating or arbitrating in communications.
In another study with $20$ designers, Wang~\etal~\cite{Wang2023} demonstrates the merits of AI generated images for early-stage ideation.
Recent works such as GenQuery~\cite{Son2023GenQuery} and DesignAID~\cite{Cai2023} demonstrate how GenAI could be useful for early-stage ideation during 2D graphic design workflow.
Similarly applied to the ideation stage, CreativeConnect~\cite{Choi2024} further shows how GenAI can may support reference images recombination.
Beyond visual design, BlendScape~\cite{Rajaram2024BlendShape} demonstrates how text-to-image GenAI and emergent VLFMs can be used to customize the video conferencing environment by dynamically changing background and speaker thumbnail layout.

As for 3D design, 3DALL-E~\cite{Liu2023} introduces a new plugin for Fusion $360$ that uses text-to-image GenAI for early-stage ideation of 3D design workflow. 
LumiMood~\cite{Oh2024} shows an AI-driven Unity tool that can automatically adjusts lighting and post-processing to create moods for 3D scenes.
Lee~\etal~\cite{Lee2024} focused on 3D GenAI with different input modalities, and revealed that the prompts can be useful for stimulating initial ideation, whereas multimodal input like sketches play a crucial role in embodying design ideas.
Blender Copilot~\cite{BlenderCopilot} is a Blender plugin that enables designers to easily generate textures and materials by text.
Vizcom~\cite{Vizcom} introduces an early-stage ideation tool for automotive designers that leverages ControlNet to convert designers' sketches into reference images.
In terms of 3D design review, ShowMotion~\cite{Burtnyk2006} demonstrates how a possibly optimal views could be searched from a set of pre-recorded shots by selecting a specific 3D element in the scene.
However, the closed-form searching algorithm only considered the $L2$-distances of each shots to the selected target, neglecting the textual feedback comments.

Inspired by these works, \sysname ~ demonstrates how to integrate text-to-image GenAI models into the 3D design feedback creation workflow.
Unlike early-stage \emph{ideation}, \emph{modeling}, or \emph{image editing} tools, \sysname~ is a \emph{review} tool that aims to enable feedback providers who might not have professional 3D and image editing skills to easily create companion images for the textual comments, contextualized on the viewpoints of initial 3D design.
\sysname's overarching tenet is to help feedback providers focus on text typing --- the primary tasks while creating design feedback.

%% file: 3-formative-study.tex
\section{Formative Studies}\label{sec::formative}
We conducted two formative studies to understand the design considerations.
Specifically, we first interviewed with professional 3D designers to understand current practices and the challenges of creating companion images for 3D design review (Sec.~\ref{sec::formative::needs}). 
We then analyzed design feedback from an online forum to understand unique 3D design characteristics that feedback writers want to convey (Sec.~\ref{sec::formative::online}). 
This process resulted in three design considerations (Sec.~\ref{sec::formative::dc}).

\subsection{Formative Study 1: Preliminary Needfinding Study}\label{sec::formative::needs}

The first study aims to understand the current professional review practices and pain-points in creating companion reference images. 
Professional 3D designers were recruited due to their extensive experience as \emph{both} 3D designers and feedback providers.

\vspace{+4px}
\noindent{\bf Participants.}
We recruited FP1 and FP2 as the \textbf{F}ormative study \textbf{P}articipants from \formativeStudyRecruitingSource.
FP1 is a modeler and digital design lead, with approximately $20$ \textbf{Y}ears \textbf{o}f \textbf{E}xperience (YoE) of 3D design and $12$ YoE in professional 3D design review process.
FP1 also has extensive experience for 3D game design as a freelance.
FP2 is currently a senior designers specialized in 3D texture design. FP2 has approximately $23$~YoE of 3D design and $15$ YoE in 3D design review.
While both participants have strong experience on using professional 3D software for automotive design such as Autodesk VRED, they also considered themselves as experts in most generic 3D software, \incl~Blender, Adobe Substance Collections as well as Autodesk Maya and Alias.
Although FP2 has tried Midjourney~\cite{Midjourney}, both participants have not used GenAI in the professional settings.

\vspace{+4px}
\noindent{\bf Methods.}
Participants first described their demographics including past 3D designs and design review experiences. 
We then conducted semi-structured interviews with each participants, focusing on two guiding questions: {\it ``what does the typical 3D design workflow look like''} and {\it ``what are the potential pain points when creating reference images for design feedback''}. 
For the second question, we further probed participants to discuss how GenAI tools could help with this task. 
The interview was open-ended and participants were encouraged to discuss their thoughts based on their professional experience.
All interviews were conducted remotely, which were then analyzed thematically using deductive and inductive coding approach~\cite{Bingham2021} (see Fig.~\ref{fig::codebook-online-dataa-nalysis-formative-study-1} in Appendix~\ref{sec::app::codebook} for the generated codebook).
This interview on average took $41$~min with each participants.

\vspace{+4px}
\noindent{\bf Findings.} 
Overall, we identified three key findings.

\vspace{4px}\noindent$\bullet$~{\bf Creating reference images starts with finding supporting 3D camera views.}
Both participants recognized the importance of finding supporting camera views in the 3D scene to aid written feedback. This task is typically done through manual exploration of the screen and taking screenshots. \blue{For example, FP1 described:}

\begin{quote}
{\it ``Most of the time, we'll simply take screenshots. We can then markup over the screenshots. Or sometimes the managers will actually take screenshots and markup what they want, or give examples of what they want and send back. [...] I typically use Zoom to record my feedback, because in our regular design software, there is not a lot of functionality like this right now''}~(FP1). 
\end{quote}

FP1 further emphasized the importance of having these supporting 3D camera views when discussing with non-technical collaborators like managers or clients. for example: {\it ``sometimes, we will just hold an online meeting. They'll talk about like from rear view, or from the top. And we will navigate the design and show them to better understand their feedback''}.

\vspace{4px}\noindent$\bullet$~{\bf Conveying changes requires additional work on the reference images.}
When it comes to creating an effective reference image for 3D design feedback, participants reported three main approaches: scene editing, using existing example images, or GenAI.

\vspace{+4px}\noindent {\bf (1) Scene editing.}
Participants reported spending time to mock up changes directly on the collected screenshots. 
FP2 described the complexities of preparing reference images for material design: {\it ``for making reference images for interior and definitely for color materials, where they apply fabrics to the seats and the door panels and like a leather grain to the instrument panel, stuff like that, it can be a little bit involved. It's not so quick. So I usually do it quickly in Photoshop. But ideally, you can also do it in the visualization software, like the VRED. I do think preparing these reference images takes the longest.''}.

\vspace{+4px}\noindent {\bf (2) Using existing reference images.}
Some design review stakeholders might resort to using some existing photographic examples of what they want to convey. For example, FP1 mentioned that {\it ``when we get to the more advanced review with the stakeholders outside our studio, sometimes, they might simply show some other example images to demonstrate what they want. But we would generally work with them iteratively, just to make sure there is no miscommunication''}.

\vspace{+4px}\noindent{\bf (3) GenAI.}
Emerging GenAI tools are seen as a promising way to generate reference images. Both participants recognized the potential benefits for visual reference creations and creativity support that GenAI tools could provide in 3D design review workflow. A key benefit that FP1 emphasized is the low barrier of entry for non-technical users: {\it ``using texts to generate image seem to be flexible and simple for those who do not know image editing software''}. 
FP2 emphasized how images generated using GenAI technologies could inspire new ideas: {\it ``GenAI is like Pinterest on steroids. You're already using Pinterest, but with GenAI you can create even more interesting inspiration. I think you tend to see the same images once in a while on Pinterest. Because if people are picking the same type of images, and you're kind of like in this echo chamber kind of thing. I remember the first few months when I started playing with Midjourney, my brain just got kind of warmed and hot. It was getting massaged! Because I'm seeing these crazy visuals that my brain isn't used to, like these weird combinations of things. I think it was a very good way for ideation. It can be also very stimulating for feedback providers to think and create suggestions''}.

\vspace{4px}\noindent$\bullet$~{\bf Controls for image generation.}
Despite the potentials of using GenAI to generate reference images, both participants mentioned that it can be frustrating to generate the right image using \emph{only} text prompts: {\it ``it's like a slot machine. I think designers kind of occasionally like happy surprises. You do 20 and maybe one is cool. But after a while, I think it gets a little frustrating and you really want more control over the output. I know there's a lot of this control on that kind of stuff, for example helping you control a little bit more perspective and painting, and stuff like that, however it is still very hard to visualize the many feedback suggestions, for example, with a small change of a particular assets components''} ~(FP2).

\vspace{4px}\subsection{Formative Study 2: Analysis of Real-World 3D Design Feedback}\label{sec::formative::online}
The first study only allows us to understand how reference images are used in current 3D design review workflow, we still need to examine what information reviewers typically encode in their feedback, and how visual references are used to convey such information.
While our semi-structured interview offers insightful thoughts from professional designers, it is difficult to analyze real-world design feedback in an ecologically valid settings, as most of design feedback in professional settings are not publicized.
Additionally, the feedback providers may also include those beyond professional 3D designer, who might not have prerequisite skills for using 3D and image editing software.

\vspace{4px}
\noindent{\bf Data Curation.}
We collected 3D design feedback from Polycount~\cite{polycount}, one of the largest free online community for 3D professionals and hobbyists. 
Polycount allows its members to post 3D artworks and receive design feedback from the community.
Although Polycount posts tend to be centered around 3D game design, the 3D assets that are discussed broadly can include a wide variety of 3D design cases like characters, objects, and scenes design.
These assets are also common in other 3D design domains.
Therefore, the outcomes yielded from our analysis could also be generalized into broader 3D applications.
Next, We selected threads within one month starting from June 2023.
Since our focus is to understand the characteristics of real-world design feedback, we only focus on the section of ``3D Art Showcase \& Critiques''~\cite{polycountCritiques}.
Irrelevant topics such as announcements and discussions related to 3D software were excluded.
Our data curation led to $15$ discussion threads, including $99$ posts from $15$ creators and $36$ feedback providers. 
Among $15$~ threads, eight threads focus on the design of characters (\eg~human and gaming avatars), three threads focus on the design of objects, and four threads focus on the scene design.

\vspace{4px}
\noindent{\bf Analysis Methods.}
We used inductive and deductive coding approach~\cite{Bingham2021} to label each design feedback posts thematically.
We aim to understand what were the primary focus of feedback and how the feedback was externalized.
Our codebook (see Fig.~\ref{fig::codebook-online-data-analysis} in Appendix~\ref{sec::app::codebook}) was generated through five iterations.

\vspace{4px}
\noindent{\bf Findings.}
Our analysis leads to findings under five themes (Fig.~\ref{fig::codebook-online-data-analysis}). 
We use OFP\texttt{<Thread\_ID>}-\texttt{<Feedback\_Provider\_ID>} to annotate \textbf{O}nline design \textbf{F}eedback \textbf{P}roviders. 
For example, OFP1-2 indicates the \emph{second} feedback provider in the \emph{first} discussion thread.

\begin{figure*}[t]
    \centering
    \includegraphics[width=\textwidth]{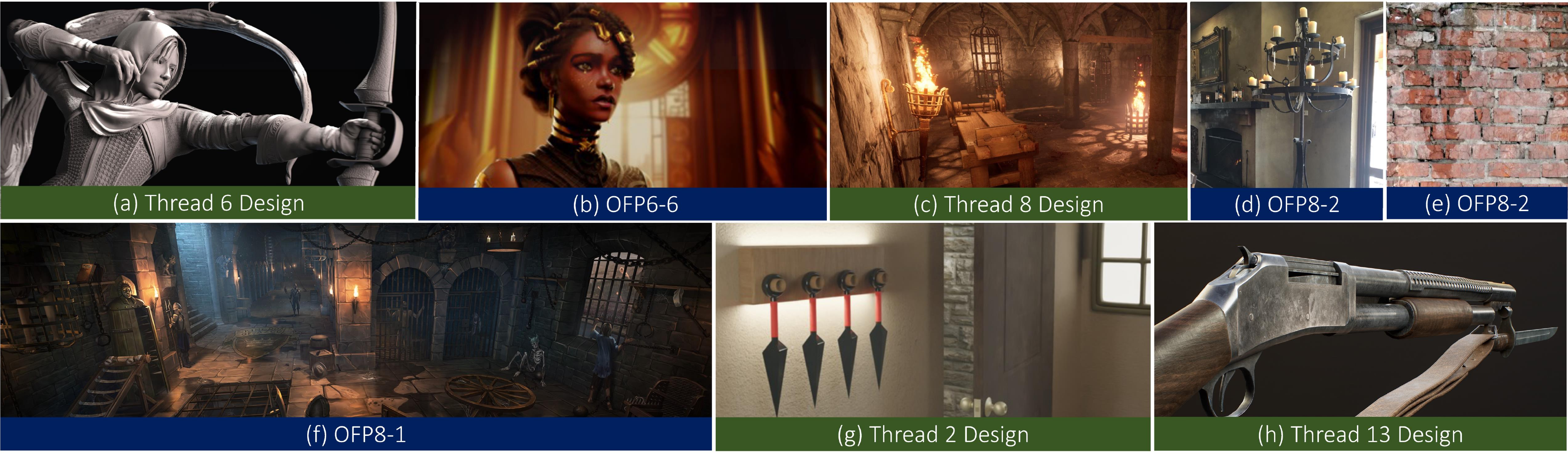}
    \caption{Example reference images from Polycount~\cite{polycount, polycountCritiques}. (a) Initial design of discussion thread 6 and (b) feedback from OFP6-6; (c) initial design of discussion thread 8 and feedback from OFP8-1 (f) and OFP8-2 (d - e); (g) initial design of discussion thread 2; (h) initial design of discussion thread 13. Green and blue labels indicate the associated reference images are from creators and feedback providers respectively.}
    \label{fig::polycount-example}
\end{figure*}

\vspace{+4px}
\noindent$\bullet$~{\bf Reference images are important to complement textual comments, but creators might need additional ``imaginations'' to transfer the gist of the visual imagery.}
One common approach for creating visual reference is to use internet-searched images. 
However, online images tend to be very different to the original design context, so feedback providers usually  write additional texts to help designers understand the gist of reference images, contextualized on the original design.
For example, while suggesting to change the color tones of the designed character (Fig.~\ref{fig::polycount-example}a), OFP6-6 used a searched figure (Fig.~\ref{fig::polycount-example}b) as a reference image and suggest: {\it ``in terms of tone, maybe look here [refer to Fig.~\ref{fig::polycount-example}b] too, just for its much softer tones across the surface''}. 
Similarly, as for scene design, OFP8-2 used Fig~\ref{fig::polycount-example}d and Fig~\ref{fig::polycount-example}e to suggest feedback for a medieval dungeon design (Fig.~\ref{fig::polycount-example}c): {\it ``the rack and some of the barrels and other assets look very new and doesn't have the same amount of wear or appear to have even been used. Need to think of the context and the aging of assets that would have similar levels of wear or damage [...] [for the wall design] I'd probably suggest to use more variation/decals, with some areas of wet or slight moss, or cobwebs''}.
OFP8-1 also similarly found another internet searched image shown in Fig.~\ref{fig::polycount-example}f as the example to suggest the design of shadow and color tone: {\it ``right now the shadows are way too dark, almost completely black which makes it really hard to tell what is going on, and causes losing a lot of detail in some parts [...] I would look into adding in some cool colored elements into your environment to help balance the hot orange lighting you have. Here is an example with less extreme lighting''}.
However, the reference images attached by feedback providers were not contextualized on the initial design (\cf~Fig.~\ref{fig::polycount-example}b \vs~Fig.~\ref{fig::polycount-example}a and \cf Fig.~\ref{fig::polycount-example}d - f \vs~Fig.~\ref{fig::polycount-example}c), and could possibly lead to confusions for the creators.

\vspace{+4px}
\noindent$\bullet$~{\bf The suggestions conveyed by the design feedback can be either the revision of specific part(s) or the redesign of entire assets.}
While much feedback such as Fig.~\ref{fig::polycount-example}d - f focused on the major revision (or even redesign) of the entire environment, we found much design feedback only emphasize the change of specific part(s) of the design assets, while keeping the rest of design remains unchanged.
For example, for Fig.~\ref{fig::polycount-example}g, OFP2-1 suggested only the change of the knives without commenting on the other part of asset: {\it ``the knife storage seems a bit unpractical. As they are knives, one would probably like to grab them by the handle? Maybe use a belt instead, with the blades going in, or some knife block as can be found in some kitchens''}.
Similarly, in the design of Fig.~\ref{fig::polycount-example}h, OFP13-3 suggested the additional modelling while satisfying the rest of the design: {\it ``looks cool! With close up shots it would be nice if bolts/screws were modeled''}.
Although such intended changes are often involved with specific part(s) of the assets, some design feedback explicitly request designers to ``visualize'' how the assets might be integrated into a different environment. 
These feedback help inspire creators to better polish their artwork.
For example, while providing feedback for polishing the tip and edge of the gun in Fig.~\ref{fig::polycount-example}h, OFP13-2 asked the creator to {\it ``just think about how this gun is used, and where wear and damage would be''}.

\vspace{+4px}
\noindent$\bullet$~{\bf Although some design feedback provides actionable solutions, others offer potential exploratory directions.}
Despite some design feedback contains specific and actionable steps, many critiques only suggested the problems, without offering suggestions on how to address the problem.
For example, for an ogre character design, OFP1-4 suggested: {\it ``I think it could use some better material separation. The cloth and the metal (maybe even the skin too) seem to have all the same kind of noise throughout''}.
Some design feedback may indicate the potential exploratory directions that might require the creators to explore by trial and error.
For example, while designing a demon character, OFP3-2 wrote: {\it ``I think that the shape is too round and baroque. More angular form should work better''}.

\subsection{Summary of Design Considerations}\label{sec::formative::dc}
We highlighted the significance of using reference images in 3D design feedback, applicable to both synchronous and asynchronous design review sessions discussed in Sec.~\ref{sec::formative::needs} and Sec.~\ref{sec::formative::online} respectively.
Both studies showed that the process of creating good reference images still remains difficult. 
These challenges come from the complexities of 3D design, requiring feedback providers to be well-versed in 3D skills to describe changes efficiently. 
While emerging GenAI are deemed promising to synthesize reference images for 3D design, we have seen from FP2's testimonies that using such tools directly without dedicated interface support can hardly meet the needs of 3D design feedback visualizations.
To integrate text-to-image-based GenAI into 3D design feedback creation workflow, we propose three key \textbf{D}esign \textbf{C}onsiderations (DCs):

\vspace{4px}
\noindent {\bf DC1: Design controlled visualization for both local and global changes.}
Our second formative study showed the importance of having a {\it controlled} way to synthesize reference images for both local (\ie~only specific part(s) of the asset need to be changed while keeping the remaining part consistent with current design) and global changes (\ie~major revision or even redesign of the most parts of the artwork).
Demonstrated in Fig.~\ref{fig::polycount-example} and discussed by FP2, although searching for images is commonly used, finding an exact or sufficiently similar design online is difficult. To address this, feedback providers often have to write elaborated and detailed comments explaining the differences and areas of focus.
While vanilla text-to-image GenAI methods offer powerful tools for image generation, the resulting images are usually not contextualized in the initial design. 
Therefore, it is critical to develop a novel interface that could leverage the power of Gen AI to generate in-context imagery at various scales seamlessly integrating this into 3D design.

\vspace{4px}
\noindent {\bf DC2: Provide ideation and creativity support for exploratory design feedback.}
We learned that feedback providers often would like to use the reference image as a source of inspiration, as it enables them to think about the alternative ways to  improve to the 3D design.
For example, some design feedback, like OFP13-2, requested the creators to imagine how the artwork would look like after being integrated into a bigger environment. 
Therefore, reference images should also demonstrate how the 3D model might look like when used in difference scenarios, inspiring both creators and feedback providers.

\vspace{4px}
\noindent {\bf DC3: Offer ways for feedback providers, including those without 3D and image editing skills to accompany their comments with meaningful in-context visualizations.}
Feedback providers does not have to possess design skills, as shown in Formative Study 1. Thus, feedback providers may also encompass individuals without professional 3D or image editing skills, such as the clients.
Without these skills, one can spend tremendous effort on even the simplest tasks, such as selecting a good view by navigating the scene with an orbit camera, inserting a novel object, or changing a texture or color of an existing object~\cite{CC3DModel2023}.
Therefore, the interface design should provide simple ways for feedback providers to navigate the 3D scene and create reference images without 3D and image editing skills.

%% file: 4-system.tex
\section{Memo-Vis}\label{sec::system}
Overall, \sysname~includes a {\it 3D explorer} (Fig.~\ref{fig::teaser}a), where feedback providers can explore different viewpoints of the 3D model, and a {\it rich-text editor} (Fig.~\ref{fig::teaser}b) where the textual feedback is typed.
Unlike image editing (\eg~Photoshop~\cite{Photoshop, GenFill2023}) and ideation tools (\eg~Vizcom~\cite{Vizcom}), \sysname~is a {\it review} tool for 3D design.
The key tenet is to enable feedback providers to focus on {\it text typing} --- the primary task for creating 3D design feedback --- while using AI-augmented rich-text editor to create companion images that illustrate the text.
As a {\it review} tool, \sysname ~ aims to enable feedback providers to create reference images that could efficiently convey the gist of the textual comments, instead of images that are visually aesthetic.
\sysname's main features include a viewpoint suggestion system and three image modification tools.

\subsection{Real-Time Viewpoint Suggestions}\label{sec::system::view_suggestion}
In \sysname, users control an orbit viewing camera with $6$DoF to render the 3D model on screen space similar to mainstream 3D software. 
However, navigating and exploring a 3D scene with a mouse can be tedious for users lacking experience in 3D software.
To enable feedback providers without 3D and image editing skills more efficiently create in-context visualizations for the textual feedback (\textbf{DC3}),
\sysname~automatically recommends viewpoints as the feedback provider continuously typing feedback.
Fig.~\ref{fig::teaser}c shows an example where a closer (and possibly better) viewpoints, shown as thumbnails, of the standing desk are recommended. 
The viewpoint being selected will be anchored with the textual comment and instantly reflected on the 3D explorer.
Feedback providers can ignore the suggestions when the suggested views are less helpful.

\begin{figure*}[t]
    \includegraphics[width=\textwidth]{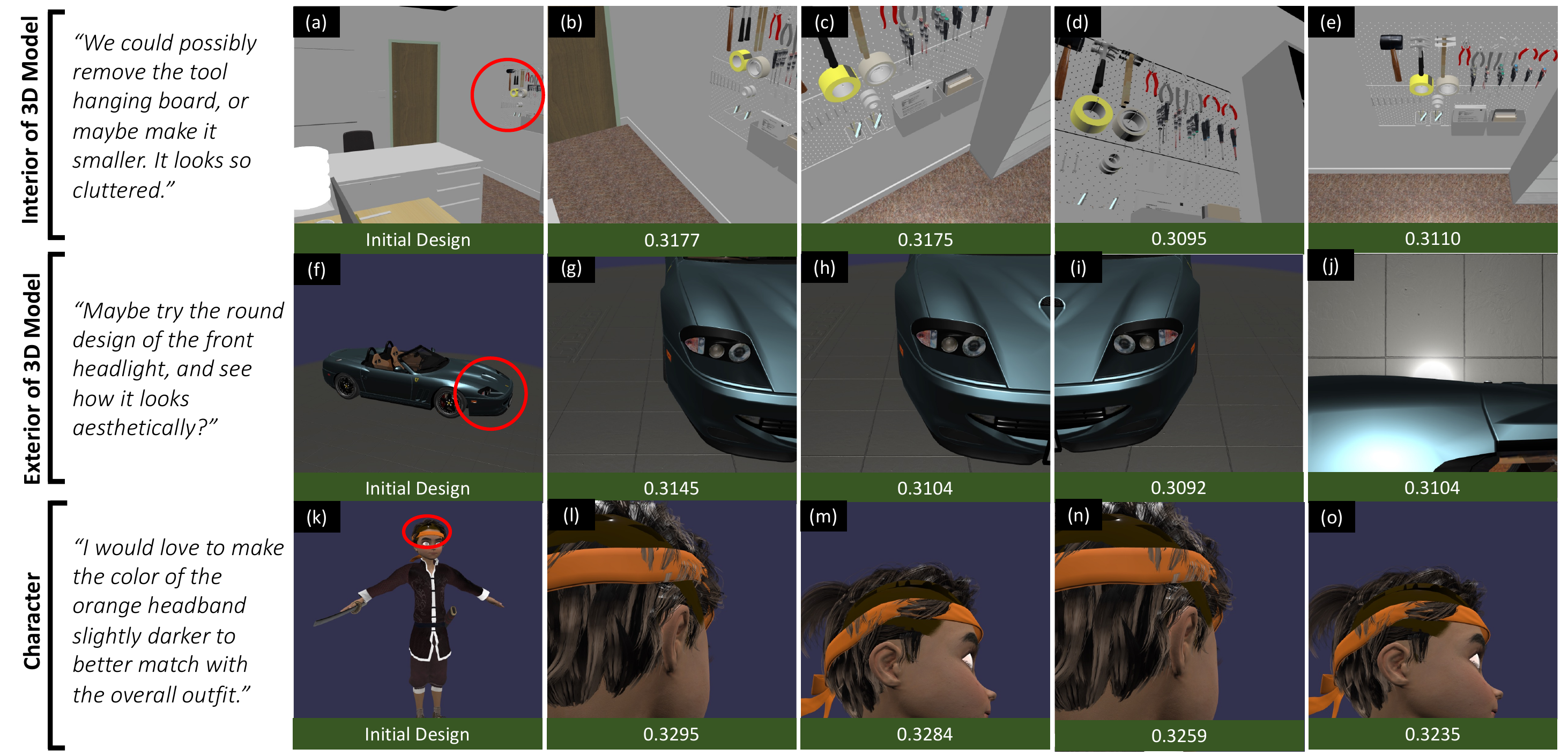}
    \caption{Examples of the suggested viewpoints based on the typed feedback comments (leftmost column). We show viewpoints with top-$\bm{4}$ highest CLIP similarity scores for an office 3D model (a - e), a car model (f - j), and a samurai boy model (k - o). a, f, and k show the bird-eye view of the initial 3D model where red circle highlight the focus of textual comments. The cosine similarity scores are shown at the bottom of each suggested viewpoints (b - e, g - j, l - o).}
    \label{fig::clip-demo}
\end{figure*}

To achieve this, we use the pre-trained CLIP model~\cite{CLIP} to find viewpoints with highest cosine similarities to the textual feedback.
Indeed, CLIP is trained on $\sim 400$~millions text-image pairs with a constrastive loss~\cite{Chopra2005ContrastiveLoss}. 
The system benefits from the human biases in image acquisition~\cite{Hentschel2022, Voigt2023Paparazzi}. For example, despite the vagueness of the text, \eg~{``office desk''}, the pre-trained CLIP model would score the desk from front and top view higher than that from the back and bottom view, as it is more common to take front-facing pictures of desks.
Formally, we parameterize the orbit camera with six parameters: the 3D point that the camera is looking at $(t_x, t_y, t_z)$ as well as its distance $r$ to the camera, and the longitudinal and latitudinal rotation $\alpha$, $\beta$. Each viewpoint can thus be represented by the tuple $\bm{v} = (\alpha, \beta, r, t_x, t_y, t_z)$, with $\alpha \in [0, \pi]$ and $\beta \in [0, 2\pi]$.
Our goal is to search $\hat{\bm{v}} = \mathbf{argmax}_{\bm{v} \in \bm{V}} cos\{f_{text}(\bm{t}) , f_{image}(\bm{I}_{\bm{v}})\}$, where $f_{text}(\cdot)$ and $f_{image}(\cdot)$ represent the CLIP encoding for text ($\bm{t}$) and screen space image ($\bm{I}_{\bm{v}}$) that is associated with a specific viewpoint $\bm{v}$.
During pre-processing, we sample multiple possible viewpoints and encode their corresponding renderings via CLIP to create a database of viewpoints. 
At inference time, we encode the textual comment via CLIP and perform a nearest-neighbor query in the database, which can be done without breaking the interactive flow.

\vspace{+4px}
\noindent$\bullet$~{\bf Pre-processing}. 
We first compute the bounding box of the 3D model.
We then discretize the $x$-, $y$-, and $z$-axis into five bins, leading to $5^3 = 125$ sampled target position $(t_x, t_y, t_z)$.
Similarly, we sample $\alpha$ and $\beta$ with $30^{\circ}$ intervals into $12 \times 6 = 72$ possibilities.
We sample $r$ from $\{0.5, 1.0, 1.5\}$ to create \emph{close}, \emph{medium}, and \emph{far} views.
For each 3D model, this pre-processing phase takes around $3$ - $5$ minutes, and results in a $\bm{D} \in \mathbb{R}^{27K \times 500}$ matrix, including $27$~k view points, each encoded via CLIP into a $500$~ dimensional feature vector. 

\vspace{+4px}
\noindent$\bullet$~{\bf Real-time inference}. 
As the feedback provider typing, the textual comment ($\bm{t}$) is encoded, and the top-$4$ nearest views under cosine similarity are retrieved. The feature runs every time the user stops typing for $500$~ms and takes under a second to compute.

\vspace{4px}
Fig.~\ref{fig::clip-demo} shows examples of suggested viewpoints of the pegboard in an office (Fig.~\ref{fig::clip-demo}a - e, as an interior design example),  the headlight of a car (Fig.~\ref{fig::clip-demo}f - j, as an exterior design example), and the headband of a samurai boy (Fig.~\ref{fig::clip-demo}k - o, as a character design example).
Although \sysname~ provides real-time viewpoint suggestions, it still allows feedback providers to manually navigate the view. For instance, they can manually find the view before writing textual comments or adjust the view based on the suggestions.

\subsection{Creating Reference Images with Rapid Image Modifiers}\label{sec::system::modifiers}
Guided by \textbf{DC2}, providing ideation and creativity support are crucial for creating design feedback, \sysname ~ therefore uses the recent text-to-image GenAI to create reference images.
However, the generated reference images should match both textual comments and current 3D design.
Critically, \sysname~ must be able to generate images with local modification of the scene if the feedback is targeted at a specific part, or images with global edits when the feedback is a global redesign of the scene.
This design goal aims to address \textbf{DC1}, emphasizing on the controlled visualization of the textual feedback for both local and global changes.
\sysname~ introduces three image {\it modifiers}, which operate on {\it rapid design layers}.
Feedback providers can use one or multiple {\it modifiers}~ (Fig.~\ref{fig::teaser}a) to easily compose and create reference image, on the associated {\it rapid design layer}, rendered from a selected viewpoint.
We now describe our system design and demonstrate how the {\it modifier} might interact with each {\it rapid design layers}.

\vspace{4px}
\noindent{\bf Modifier 1: Text $+$ Scribble Modifier with Scribble Design Layer}

\noindent We leverage ControlNet~\cite{Zhang2023ControlNet, Zhao2023uni} for two scenarios.
In the simpler case, the feedback is just a global texture edit on the scene that does not suggest any geometry modification. In this case,  we use a depth-conditioned ControlNet~\cite{Zhang2023ControlNet} to generate an image. 
The depth guidance ensures that the generated image is anchored in the current design, while the textual prompt generates an image that matches the feedback. 

\begin{figure*}[t]
    \centering
    \includegraphics[width=\textwidth]{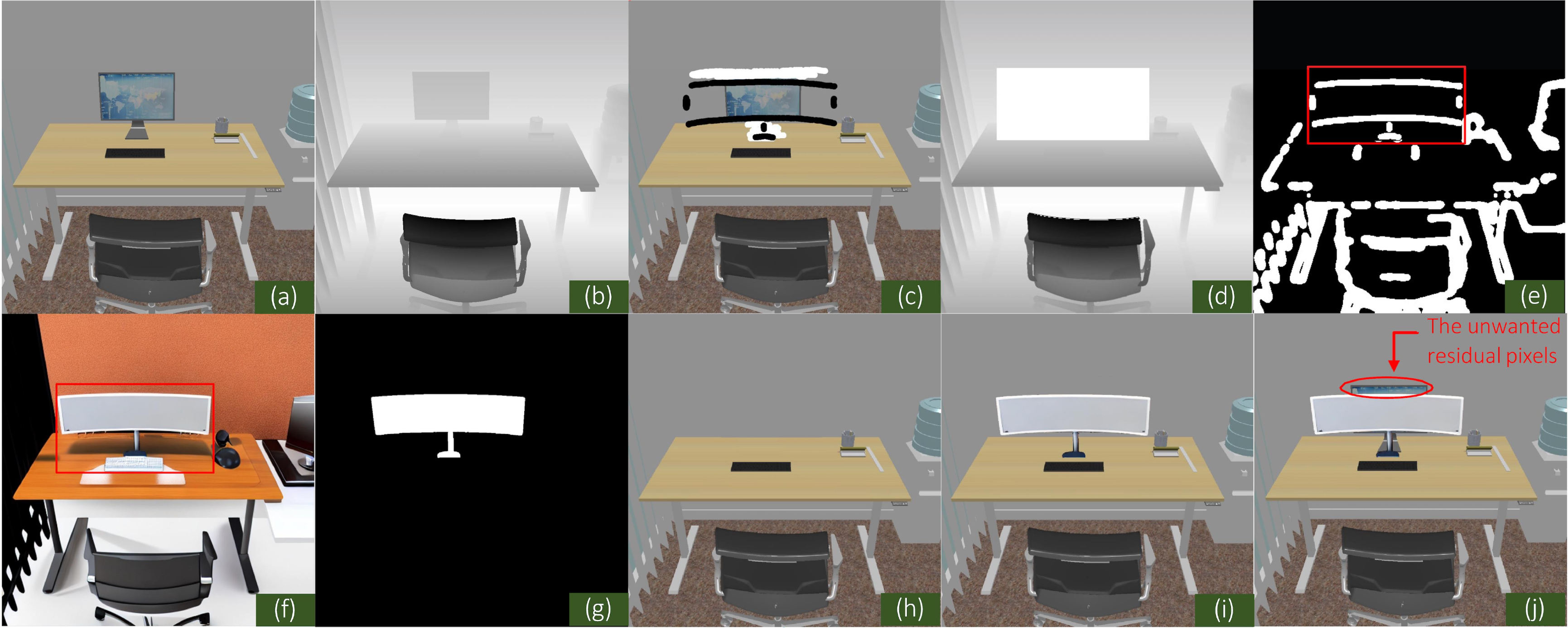}
    \caption{Examples of creating reference image using the text + scribble modifier. (a) Initial design; (b) associated depth map; (c) manually drawn scribbles (black strokes) with the white strokes indicating the removed geometries; (d) the depth map with the scribbling area being reset; (e) an aggregated scribble from the initial image and the manually drawn scribbles, where the red bounding box shows the scribbling area by feedback providers; (f) synthesized image by ControlNet conditioned by scribble $+$ depth, where the red bounding box shows the area that the feedback providers scribbled; (g) segmented mask generated by SAM; (h) initial design with the primitives describing existing computer display being removed; (i) final composed reference image; (j) final composed reference image without removing objects marked for removal by scribbling.}
    \label{fig::controlnet-scribble-demo}
\end{figure*}

If the feedback suggests to modify the geometry of part of the scene (\eg~the review from OFP2-1 for the 3D design of Fig.~\ref{fig::polycount-example}g), directly editing the depth image~\cite{Dukor2022} is impractical for feedback providers without graphic knowledge. 
Instead, we leverage a depth- and scribble-conditioned ControlNet in a somewhat more involved strategy\footnote{An inference example using depth- and scribble-conditioned ControlNet could be referred to Fig.~\ref{fig::controlnet-scribble-depth-compare} in Appendix~\ref{sec::app::compare_controlnet_depth_scribble}.}~\cite{Zhang2023ControlNet, Zhao2023uni}. 
For example, let's say the design feedback is to replace the computer display with a curved one in the scene depicted in Fig.~\ref{fig::controlnet-scribble-demo}a with the depth map shown in Fig.~\ref{fig::controlnet-scribble-demo}b, the feedback provider only has to scribble the rough shape of the new curved computer display to be added, as in~Fig.~\ref{fig::controlnet-scribble-demo}c, and provide a text prompt. 
\sysname~ starts by creating the input conditions to ControlNet. It extracts the scribble from the initial design (Fig.~\ref{fig::controlnet-scribble-demo}a) using \textbf{H}olistically-nested \textbf{E}dge \textbf{D}etection (HED)~\cite{Xie2015HED} and aggregated it with the manually-drawn scribbles (Fig.~\ref{fig::controlnet-scribble-demo}e). 
The depth map (Fig.~\ref{fig::controlnet-scribble-demo}d), aggregated scribbles and text prompt, are then fed to ControlNet to generate an image~(Fig.~\ref{fig::controlnet-scribble-demo}f).
Empirically, we set the strengths for scribble to $0.7$ and depth condition to $0.3$ emphasizing the higher importance of the newly added object(s).

In addition to changing the geometry, the generated image might modify the texture of the current design which is undesirable. To address this, we leverage automatic segmentation techniques to merge the generated object from the generated image $\bm{I}_{syn}$ \ie~the {``the curved computer display''}, back into the original render $\bm{I}_{init}$.
To achieve this, we compute the bounding box of the user scribbles (red box in Fig.~\ref{fig::controlnet-scribble-demo}e) and find the most salient object within the bounding box using SAM~\cite{Kirillov2023SAM}, leading to a segmentation mask $\bm{I}_{seg}$~(Fig.~\ref{fig::controlnet-scribble-demo}g).
\sysname~then creates the final reference image shown in Fig.~\ref{fig::controlnet-scribble-demo}j by composition:  $\bm{I}_{syn} \odot \bm{I}_{seg} + \bm{I}_{init} \odot (\mathbf{1} - \bm{I}_{seg}) $, where $\odot$ indicates broadcasting and element-wise multiplication.

This approach allows to visualize the newly added objects, but part of the initial object, \ie~the current display can remain visible, leading to unpleasing visual artifacts, circled in red in Fig.~\ref{fig::controlnet-scribble-demo}j.
To address this, \sysname~ detects the mesh primitives to be removed from the image bounding box and depth map, and re-render $\bm{I}^{\prime}_{init}$, which the same image as $\bm{I}_{init}$ but without the object to be removed. Replacing $\bm{I}_{init}$ by $\bm{I}^{\prime}_{init}$ in the composition leads to the final results, displayed to the user, and visualized in Fig.~\ref{fig::controlnet-scribble-demo}i.
Algo.~\ref{alg::remove_hidden_obj} in the Appendix~\ref{sec::app::algo} recaps the algorithm.

\vspace{4px}
\noindent{\bf Modifier 2 : Grab'n Go Modifier with GenAI Design Layer}

\begin{figure*}[t]
    \centering
    \includegraphics[width=\textwidth]{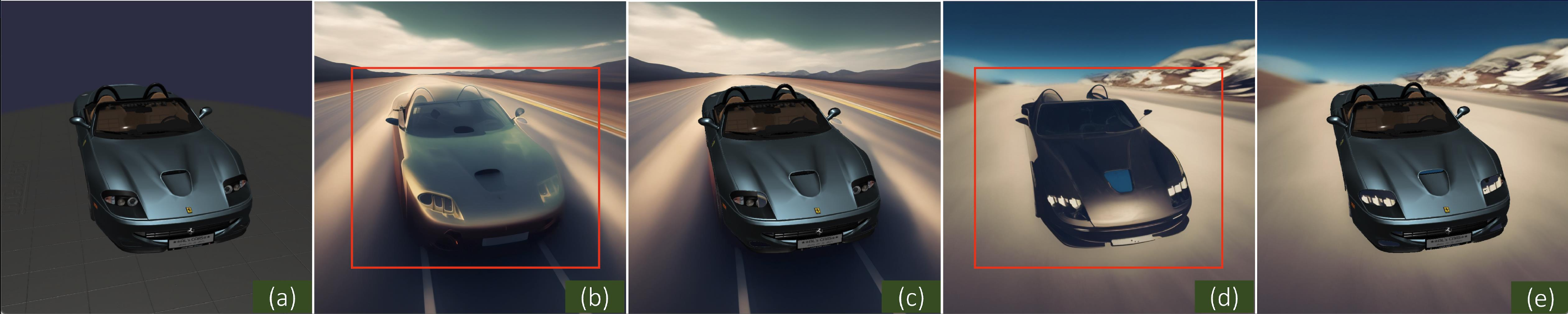}
    \caption{Examples showing how the feedback providers can stage the 3D model into different scene with the grab'n go modifier. (a) The initial 3D design of a car; (b, d) synthesized image generated by scribble + text modifier with ControlNet conditioned on depth. The prompt {``a Ferrari car driving on the highway''} and {``a Ferrari car driving on a dessert''} were used to synthesize (b) and (d), respectively. The red bounding boxes show the areas drawn by feedback providers; (c) final composed image by bringing initial design into the scene of (b); (e) final composed image by bringing initial design into the scene of (d). }
    \label{fig::staging}
\end{figure*}

\noindent The {\it grab'n go modifier} is an easy selection tools that allows the user to compose an object from the rendered image into a generated image. 
For instance, considering the car in Fig~\ref{fig::staging}a, we can generate an image of this car staged in various backgrounds (Fig~\ref{fig::staging}b and Fig~\ref{fig::staging}d) using the depth-conditioned ControlNet~\cite{Zhang2023ControlNet}. 
However, the car might have undesirable texture variations. The feedback providers can simply draw red rectangles to select the car object and replace it with the exact car in the current design, thus staging it in the desired environment.

\begin{figure*}[t]
    \centering
    \includegraphics[width=\textwidth]{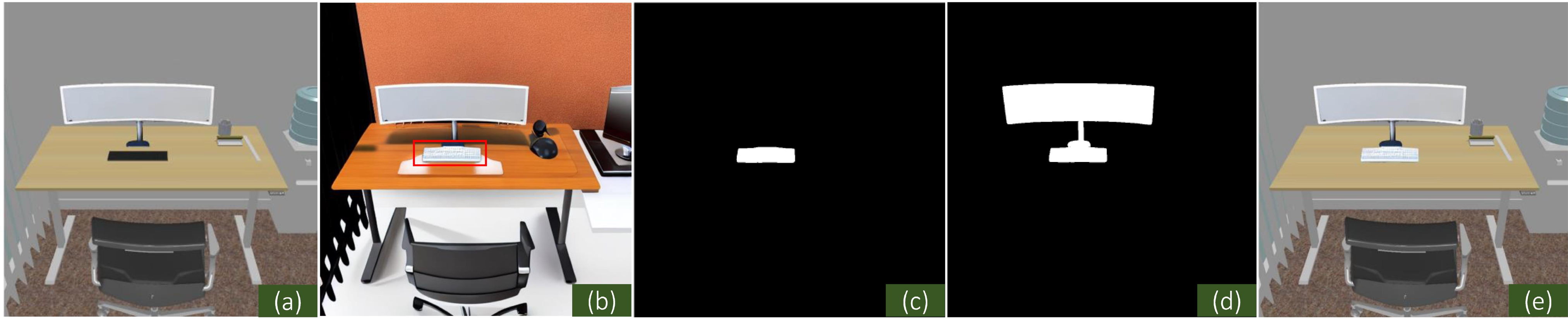}
    \caption{Examples of continuous composing. (a) Reference image of Fig.~\ref{fig::controlnet-scribble-demo}i; (b) feedback provider can draw a bounding box to indicate their intention to add the white keyboard design into the reference image; (c) segmented mask generated by SAM; (d) segmented mask by compute the union of (c) and Fig.~\ref{fig::controlnet-scribble-demo}g; (e) final reference image after including the white keyboard.}
    \label{fig::controlnet-scribble-continuous-composition}
\end{figure*}

The grab'n go modifier can also be used to do the reverse, \ie~ composing objects from the generated images into the current design. For example, if the feedback providers wants to also include the generated keyboard in our previous example with the curved display in Fig.~\ref{fig::controlnet-scribble-continuous-composition}b, they can simply draw a crop on the GenAI design layer. 
To achieve this, similar to the first modifier, we give the bounding box drawn by the user as input to SAM~\cite{Kirillov2023SAM}, and select the highest scored region as the output~(Fig.~\ref{fig::controlnet-scribble-continuous-composition}c). \sysname~then computes a final segmentation mask by \textit{union}-ing Fig.~\ref{fig::controlnet-scribble-continuous-composition}c and Fig.~\ref{fig::controlnet-scribble-demo}g that would be used to create final reference image (Fig.~\ref{fig::controlnet-scribble-continuous-composition}e). 
Formally, the final segmentation mask after $i$ times of applying {grab'n go modifier} could be computed by $\bm{I}_{seg} = \bm{I}_{seg_{i - 1}} \cup \bm{I}_{seg_i}$.
Note how this is significantly simpler than professional image editing software which commonly use the \emph{Lasso} tool~\cite{LassoPhotoshop}. 
As suggested by \textbf{DC3} emphasizing the needs for users without image editing skills, the interactions around grab'n go modifier enables feedback providers without professional image editing skills to efficiently create companion images.

\vspace{4px}
\noindent{\bf Modifier 3 : Text + Paint Modifier with Painting Design Layer}

\begin{figure*}[t]
    \centering
    \includegraphics[width=\textwidth]{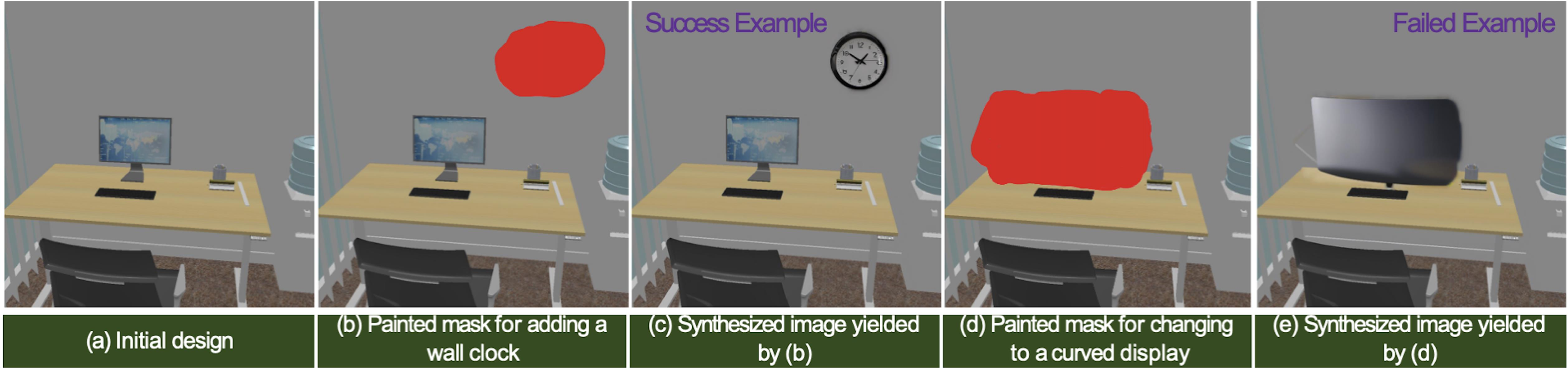}
    \caption{Examples of visualizing feedback that is related to adding new additional object with text + paint modifier. (b - c) shows a wall clock is successfully added to the view. (d) aims to add a curved computer display which is more complicated in terms of shapes, geometries and orientations. (e) demonstrates a failed attempt with {text + paint} modifier.}
    \label{fig::controlnet-inpaint-scribble-compare}
\end{figure*}

\noindent \sysname~ integrates Stable Diffusion Inpainting~\cite{Rombach2022SD} as a {\it text + paint modifier}. The user paints a selection mask on the canvas, provides a text prompt and the model generates an image. This can be used to add simple objects; remove objects; or rapidly fix the glitches caused by {text + scribble} and {grab'n go} modifiers.
This tool complements the {text + scribble modifier}.
Fig.~\ref{fig::controlnet-inpaint-scribble-compare} shows an example of adding a wall clock with {text + paint} modifier (Fig.~\ref{fig::controlnet-inpaint-scribble-compare}a - c). 
When trying to add a curved computer display, {text + paint} modifier fails to convey the intention of the design feedback (see Fig.~\ref{fig::controlnet-inpaint-scribble-compare}d). We thus argue for the necessity of user scribbling to add more complicated object, which would be hard to describe in details with text.

\begin{figure*}
    \centering
    \includegraphics[width=\textwidth]{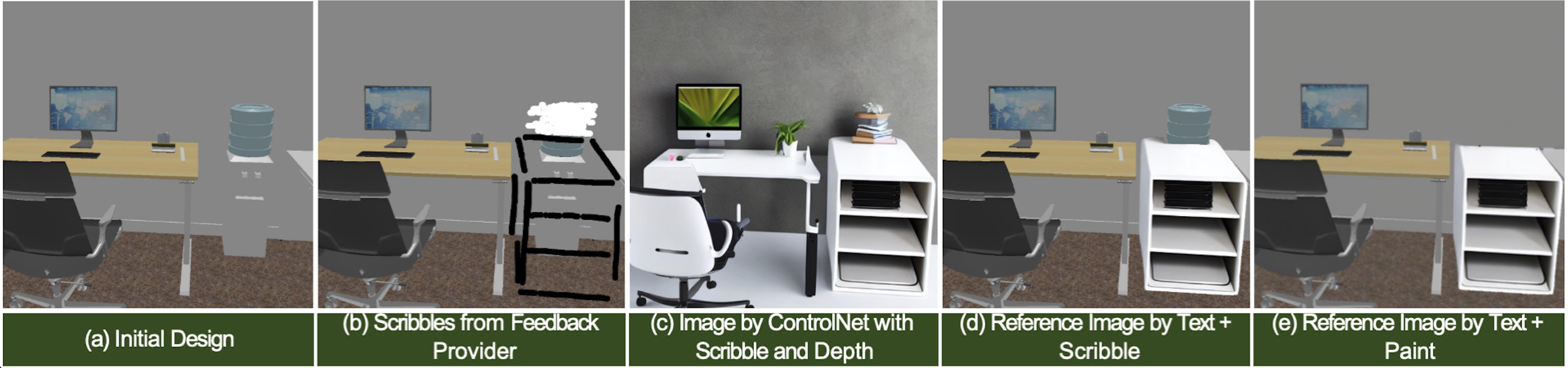}
    \caption{Text + paint modifier can be used to fix the glitches of the images created by text + scribble and grab'n go modifiers.}
    \label{fig::controlnet-inpaint-fix}
\end{figure*}

\sysname~ can also fix artifacts in {text $+$ scribble} and {grab'n go modifiers}.
Those artifact occasionally arises in Algo.~\ref{alg::remove_hidden_obj}, when the residual area occupies more than $70\%$ of the areas of the corresponding meshes (\ie~$r > r_{th}$).
For example, to change the water dispenser in an office to a storage drawer (Fig.~\ref{fig::controlnet-inpaint-fix}a), the feedback provider might leverage the {text + scribble} modifier to roughly draw the shape of a drawer (Fig.~\ref{fig::controlnet-inpaint-fix}b), leading to the drawer being extracted from synthesized image (Fig.~\ref{fig::controlnet-inpaint-fix}c) and added to the initial design (Fig.~\ref{fig::controlnet-inpaint-fix}d).
While Fig.~\ref{fig::controlnet-inpaint-fix}d may convey the ideas of adding a storage drawer next to the standing desk, the water dispenser is misleading and needs to be removed.
Feedback provider can continuously use the {text + paint modifier} to easily remove the residual areas of the top of water dispenser, leading to a more faithful final reference image (Fig.~\ref{fig::controlnet-inpaint-fix}e).

\subsection{Interactions and Implementations}\label{sec::system::interaction_implementations}
Fig.~\ref{fig::teaser} shows the design of \sysname.
With the text $+$ scribble modifier, the feedback provider scribbles while holding the mouse click.
The feedback provider can use the left and right mouse buttons to indicate the geometry of new objects and the unwanted areas, respectively.
With grab'n go modifier, the feedback providers can easily draw a bounding box using either the left or the right mouse button.
The left and right mouse button indicate keeping and removing the key object(s) inside the enclosed box of the synthesized image in the reference image, respectively.
Finally, with the text $+$ paint modifier, \sysname~enables feedback providers to click and drag left and right mouse button to specify the areas for adding (or changing) and removing objects of interests, respectively.

\sysname~was prototyped as a browser-based application to reduce the needs for feedback providers to install large-scale standalone 3D software, as partial engineering efforts to address DC3.
We used Babylon.js v$6.0$~\cite{babylonjs} to implement the 3D explorer for rendering the 3D model.
A flask server was implemented to process the inference workloads, deployed on a GPU-enabled cloud server. 
Further details of the pre-trained models we used can be referred to Appendix~\ref{sec::app::models}.

%% file: 5-evaluation.tex
\section{Evaluations}\label{sec::evaluation}
We conducted two user studies to evaluate \sysname. 
The first study aims to understand the experience of feedback providers while visualizing textual comments using \sysname, whereas the second study aims to explore how the created images by \sysname~could effectively convey the gist of 3D design feedback while being consumed by designers.
We use PF\# and PD\# to index \textbf{P}articipants for acting as \textbf{F}eedback providers and \textbf{D}esigners, respectively.

\subsection{Study 1: Creating Reference Images}\label{sec::eval::create_feedback}
Our first study was structured as a {\it within-subjects design}. 
We aim to tackle two \textbf{R}esearch \textbf{Q}uestions (RQs):

\vspace{4px}
\begin{itemize}[noitemsep, topsep=0pt, leftmargin=*]
    
    \item \textbf{(RQ1)} How the real-time viewpoint suggestion could help feedback providers navigate the 3D scene while writing feedback?
    
    \item \textbf{(RQ2)} How the three types of image modifiers could help feedback providers visualize the textual comments for 3D design?
    
\end{itemize}
\vspace{4px}

\vspace{+4px}\noindent{\bf Participants.}
PF1 - PF14 (age, $M = 23.36$, $SD = 2.52$, \incl~seven males and seven females) were recruited as the feedback providers.
While all participants have experience of writing design feedback, most participants had limited experience working with 3D models. We believe that the design feedback providers do not necessary to possess design skills. Among recruited participants, only PF2, PF10, and PF14 were confident of their 3D software skills.
Most of participants did not have experience of creating prompts and using GenAI, although PF1 and PF2 considered themselves as experts of using LLM; PF13 was confident of his proficiency of using text-to-image GenAI. 
Details of participants' demographics \blue{and power analysis} can be referred to Appendix~\ref{app::evaluation_participant}.

\vspace{+4px}\noindent{\bf Interface Conditions.}
Participants were invited to use two interface \textbf{C}onditions (C1 - C2) to create and visualize textual comments:

\vspace{4px}
\begin{itemize}[noitemsep, topsep=0pt, leftmargin=*]

    \item \textbf{(C1)~Baseline.}~Participants were required to create reference image(s) using search and/or hand sketching. We aim to mock up current design review practices based on the findings from Formative Study 1. Participants were not required to use existing GenAI-based image editing tools like Photoshop~\cite{photoshopAI}; while FP2 acknowledged GenAI as a promising approach to generate reference images, its lack of control makes it impractical (\S\ref{sec::formative::needs}); unlike FP2 who are professional designers, feedback providers may not possess skills on using image editing software. Participants were instructed to use their preferred search engine for finding well-matched images. PowerPoint was optionally used for sketching and annotations.
    
    \item \textbf{(C2)~\sysname.}~Participants were invited to use \sysname~ to create reference images along with textual comments.
    
\end{itemize}

\vspace{4px}
\noindent{\bf Tasks.}
Each participant was instructed to complete three different design critique \textbf{T}asks (T1 - T3) with C1 and/or C2. 
To prevent learning effect, T1 - T3 used \emph{different} 3D models, created by \emph{different} creators. 
For each design critique task, participants were instructed to create at least two design comments, with each comments being accompanied by at least one reference images.
We used T1 to help participants get familiar with C1 and C2, with which participants were instructed to critique a character model of a samurai boy. 
All data collected from T1 was excluded from all analysis. 
For T2 and T3, participants were asked to make the bedroom and the car more comfortable to live in and drive with, respectively.
While T2 and T3 used different 3D models, the skills needed to create design feedback are the same.
Details of the study tasks could be referred to Fig.~\ref{fig::study-models} and Appendix~\ref{app::study_tasks}.

\vspace{+4px}\noindent{\bf Procedures.}
Participants were invited to complete a questionnaire to report demographics, 3D software, 3D design, and GenAI experiences.
They were then introduced to \sysname, and were given sufficient time to learn and get familiar with both C1 and C2 while completing T1.
For those without prompt writing experience for GenAI, sufficient time was provided to go through examples on Lexica~\cite{lexicaart} and practice using FireFly~\cite{AdobeFirefly}.
Upon feeling comfortable with C1 and C2, participants were invited to complete T2 and T3.
To prevent the sequencing effects, we counterbalanced the order of interface conditions.
Specifically, PF1 - PF7 were required to complete T2 with C2, followed by completing T3 with C1.
Whereas, PF8 - PF14 were required to complete T2 with C1, followed by complete T3 with C2.
Comparing feedback creations for T2 and T3 were out of our scope, we therefore did not counterbalance the order of the tasks.
After each task, participants were invited to rate their agreement of four \textbf{Q}uestions (Q1 - Q4) in a $5$-point Likert scale, with a score $>3$ being considered as a positive rating.

{\color{black}

\vspace{4px}
\begin{itemize}[noitemsep, topsep=0pt, leftmargin=*]
    
    \item \textbf{(Q1) Navigating the Viewpoint}: {\it ``it was easy to locate the viewpoint and/or target objects while creating feedback.''}
    
    \item \textbf{(Q2) Creating Reference Images}: for the task completed by \sysname~ (C2), we used the statement {\it ``it was easy to create reference images with the image modifiers for my textual comments''}. For the task completed by baseline (C1), we used the statement {\it ``it was easy to create reference images with the method(s) I chose''}.

    \item \textbf{(Q3) Explicitness of the Reference Images}: {\it ``the reference images easily and explicitly conveyed the gist of my design comments.''}

    \item \textbf{(Q4) Creativity Support}: {\it ``the reference images helped me discover more potential design problems and new ideas.''}
\end{itemize}
\vspace{4px}

}

A semi-structured interview was also conducted focusing on participants' rationales while evaluating Q1 to Q4.
The study on average took $57.34$ min ($SD = 7.72$~min).  

\vspace{+4px}\noindent{\bf Measures and Data Analysis}.
To address RQ1, we measured the {\it navigating time} for each textual comment, defined by the time that feedback providers spent while navigating and exploring the 3D model.
With the Shapiro-Wilk Test~\cite{Shapiro1965}, we verified the normal distribution of the measurements under each condition ($p > .05$).
One-way \textbf{An}alysis \textbf{o}f \textbf{Va}riance (ANOVA)~\cite{Girden1992} ($\alpha = .05$) was therefore used for statistical significance analysis.
The eta square ($\eta^2$) was used to evaluate the effect size, with $.01$, $.06$ and $.14$ being used as the empirical thresholds for {\it small}, {\it medium} and {\it large} effect size~\cite{Cohen1998}.
We used thematic analysis~\cite{Braun2012} as the interpretative qualitative data analysis approach, along with deductive and inductive coding approach~\cite{Bingham2021} to analyze participants' responses during semi-structured interviews, to better understand participants' thoughts and uncover the reasons behind the measurements and survey responses.
We used the initial codes from Q1 - Q4.
We first read the transcripts independently and identified repeating ideas using initial codes derived from Q1 to Q4. 
Next, we inductively come up with new codes and iterate on the codes as sifting through the data.
The final codebook can be found in Fig.~\ref{fig::codebook-study-1} in Appendix~\ref{sec::app::codebook}.

\begin{figure*}[t]
    \centering
    \includegraphics[width=\textwidth]{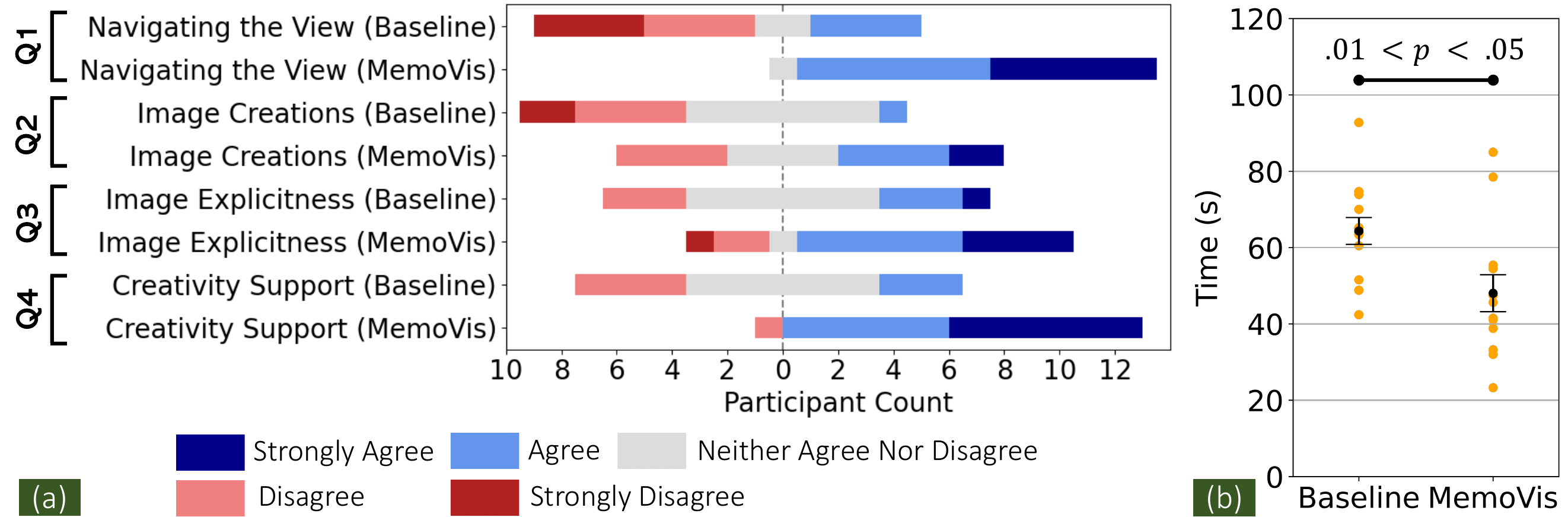}
    \caption{Survey responses and navigating time measurements from Study 1. (a) Participants' responses of Q1 - Q4; (b) the total navigating time of baseline and \sysname. PF14 was excluded from Fig.~\ref{fig::study-1-results}b, as the participant did not use the viewpoint suggestion feature.}
    \label{fig::study-1-results}
\end{figure*}

\vspace{+4px}\noindent{\bf Results and Discussions}

\noindent Most participants found our viewpoint suggestion features useful~(RQ1) and the image modifiers easy to use to visualize textual comments~(RQ2). Most participants also believed the reference images created with \sysname~could easily and explicitly convey the gist of 3D design feedback. Patterns of usages of each image modifiers could be referred to Fig.~\ref{fig::study-1-modifier-usages} in Appendix~\ref{sec::app::modifier_usages}.

\vspace{+4px}\noindent$\bullet$~{\bf How the real-time view suggestions could help feedback providers navigate inside 3D explorer (RQ1)?}
Overall, $13/14$ participants (except PF14) leveraged the viewpoint suggestion features while visualizing the textual comments.
More participants positively believed that it was easy to locate the viewpoints and target objects with \sysname, compared to the baseline~($13$~\vs $4$, Fig.~\ref{fig::study-1-results}a). 
Fig.~\ref{fig::study-1-results}b shows a significant reduction ($F_{1, 24} = 7.398$, $p = .018$, $\eta^2 = .236$) of navigating time while using \sysname~($M = 44.06$s, $SD = 18.94$s) \vs~ the baseline ($M = 66.67$s, $SD = 16.94$s).
Notably, the normality of the measurement was verified ($p_{baseline} = 0.54$, $p_{\sysname} = 0.17$).
Our qualitative analysis suggested two benefits brought by the viewpoint suggestion feature.

\vspace{+4px}
\noindent{\textbf{\textit{Providing guides to find the viewpoint contextualized on the design comments.}}}
Most participants appreciated the help and guides brought by the viewpoint suggestion feature.
For example, 
{\it ``the view angle is good that it's trying to give you a context''}~(PF1) 
and {\it ``it helped me a lot to find a nice view where I could create reference image''}~(PF2).
PF8 also suggested the potential benefits for feedback providers to make faster decision: {\it ``it helps me to make faster decisions. Sometimes I don't know which views might be better. And it gives me options that I could choose''}~(PF11).
After trying baseline, PF11 emphasized: {\it ``[without viewpoint suggestion] I have to decide on how to look. I have to make bunch of decisions in the way. It is pretty cognitively demanding''}.
PF7, who did not have prior 3D experience, initially felt {\it``confused about the directions of the viewpoint''}, while attempting to navigate the 3D scene using the 3D explorer.
After using the viewpoints suggested by \sysname, she felt {\it ``it's a much better view, as the bed could be seen from a nice view angle''}.

\vspace{+4px}
\noindent{\textbf{\textit{Locating target object(s) in a scene.}}}
Despite occasional failures and the needs of minor adjustments, most participants appreciated the benefits of being able to quickly locating target objects.
For example, PF8 acknowledged: {\it ``I feel that around 85\% of the time that the system could give me the right view that I expected, although I sometimes might still need to adjust like a zoom''}.
PF4 justified the reason of not {\it strongly agreeing} with Q1: {\it ``although it was helpful, like the system gave me a nice view suggestion. But I had to move it manually. Although that was a good starting point, I still need to make adjustments by myself''}.
During the interview, PF2 believed that {\it ``the view suggestions would be more useful for the bigger scene''}.
With the past experience of designing 3D computer games, he further commented: {\it ``sometimes I work in video games. And video game maps could become really large. And there are multiple things. For example, there's a very specific area that I find to go to, and to edit it. And then if I can type and say, for example, the boxes on the second floor of the map. And it instantly teleport me to there. Then, there is gonna save me a lot of time to find it in the hierarchy''}.
After PF14 critiqued a car design with \sysname, he commented: {\it ``I think moving around with car was easier just because it's a car instead of the room. But I believe for the bedroom it would be much helpful to have the view suggestion feature that kind of guide''}.

\vspace{4px}\noindent$\bullet$~{\bf How \sysname~could better support feedback providers to create and visualize textual comments for 3D design (RQ2)? }
Fig.~\ref{fig::study-1-results}a shows that compared to the baseline, more participants rated positively in terms of reference image creations ($6$~\vs~$1$ for Q2), explicitness of the images ($10$ \vs~$4$ for Q3), and creativity support ($13$ \vs~$3$ for Q4).
Feedback provider participants have overall used $48$~times ($42.11$\%) of text $+$ scribble modifier, $30$~times ($26.32$\%) of grab'n go modifier, and $36$~times ($31.57$\%) of text $+$ paint modifier (see Fig.~\ref{fig::study-1-modifier-usages} in Appendix~\ref{sec::app::modifier_usages} for the detailed usages of the image modifiers). 
While completing the baseline task, despite the challenges of searching well-matched internet images, participants used three main strategies to make the reference images more explicit. 
Specifically, four participants used annotations to highlight the areas that the textual feedback focus;
One participant (PF3) used two reference images to demonstrate a good and bad design, and expected the designers to understanding gist by contrasting two extreme examples;
And two participants (PF2, PF13) provided multiple reference images and hoped the designers could extract different gist from separate images.

\vspace{4px}
\noindent{\textbf{\textit{Image explicitness.}}}  
Participants appreciated the explicitness of the reference images created by \sysname, while keeping the contexts of the initial design.
For example, {\it ``I like how it \underline{keeps the context} around this picture. It could be much easier for people to understand my thoughts''}~(PF2), {\it ``it allows me to \underline{generate reference image in the same scenario}, and not some random scenario that I've pulled from the internet''}~(PF9), and {\it ``this image references are pretty easy and explicitly. It just \underline{conveys my point}. That's what matters. Now it's up to the person to make the decision to how have to make it better''}~(PF11).
Some participants like PF6 prefer to the \sysname-created reference images, compared to the searched and hand-sketched images.
PF11 highlighted the easy and convenience workflow: {\it ``this process was pretty easy. The reference image is just like pop up to me! This is something I love. Like when I provide feedback while I was teaching, I didn't explicitly tell the students like, your typography is really hard to read, you should change it to this font. But something like a bigger front, a different style, just like that''}.
In contrast, after creating feedback with baseline task, PF13 complained: {\it ``you can find tones of image on the Google. They are very realistic. They are very decorative. But it's just not related to my model, it's not in the context [...] when I use an internet image, there are more details, but there is even more confusing part. So many times, I just tweak my textual feedback, to minimize the possible confusions for the designers [...] I think if I were doing by myself, I would just spend some more time and use Photoshop to edit the internet images''}.
P14 also commented: {\it ``I typical have a specific image in mind. I think to come up with that design, [the \sysname] is much much better. When I search for something, it is typically very generic. I never search something that is very specific like, a bedroom with blue walls or something. It's easy if you're coming in with a specific design''}.

\begin{figure*}[t]
    \centering
    \includegraphics[width=\textwidth]{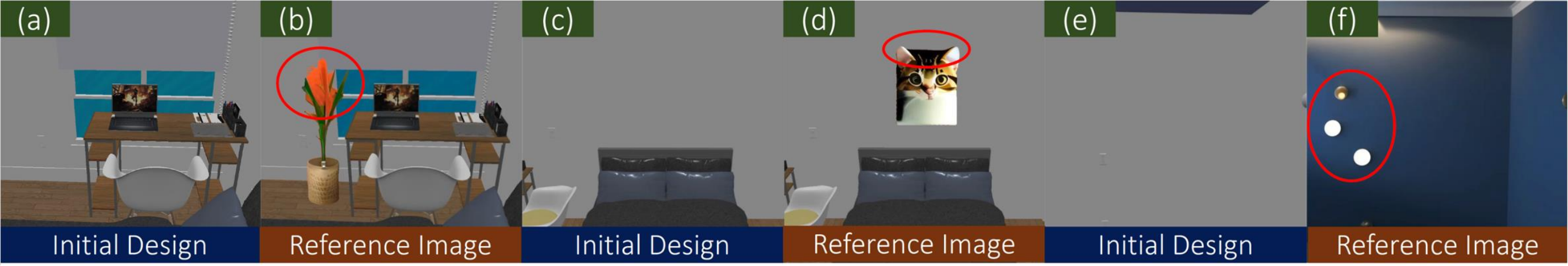}
    \caption{Examples of inspirations and creativity support. (a, c, e) show the selected viewpoints of the initial design; (b, d, f) show the created visual references by PF4. The reported unexpected components are highlighted by red circle.}
    \label{fig::study-examples}
\end{figure*}

\vspace{4px}
\noindent{\textbf{\textit{Unexpected inspirations and creativity support.}}}
Most participants recognized the benefits of \sysname~for inspiring new ideas, which is similar to FP2's comments (Sec.~\ref{sec::formative::needs}).
For example, {\it ``there are just so many possibilities. If you search through Google, your thoughts are limited by your experience. But this tool could give me so much unexpected surprises, which is a good thing and they are many times actually better than what I thought. I think it works quite well to help inspire more new ideas''}~(PF12).
PF13 particularly enjoyed the mental experience of getting inspired iteratively while refining the textual prompts: {\it ``when I was typing like the description, it was just a text like a description. I don't really have a solid image in my brain. But [\sysname] helps me shape my idea, and helps me better think iteratively [...] it's like when you're writing papers, instead of starting from the scratch, you have a draft, so it's easier to discuss and revise based on it [...]''}. 
Fig.~\ref{fig::study-examples} demonstrates three examples with unexpected creativity that were appreciated by PF4.
For example, upon seeing Fig.~\ref{fig::study-examples}b, PF4 thought out loud that: {\it `` I wasn't expecting the plant to have like the little orange leaves, but I think it was a good idea for the final design''}.

\vspace{+4px}
\noindent{\textbf{\textit{Simple and easy image creation workflow.}}}
Most participants appreciated the simple workflow for creating reference images with \sysname, and felt {\it ``very easy to use''}~ (PF14).
This was also evident by the responses for Q2 reported in Fig.~\ref{fig::study-1-results}.
Participants highlighted the merits of easy workflow facilitated by the conveniences of all image modifiers.
For example, {\it ``the system is very easy to use, although it might require a trial or two to become familiar with the modifiers. Once I understand, it is really handy and convenient to instantly create reference images that I want, which also match the design and my comments. All image modifiers are really useful, essential and indispensable!''} ~(PF1).  
PF9 particularly appreciated the option of {\it ``select and extract''} supported by grab'n go modifier: {\it ``I think the workflow is pretty good. [...] The image generation is not the only part. But I'm getting the option to select and extract [the new objects that I am trying to suggest], instead of just using some random image on Google [PF9 refers to the baseline interface condition]''}.
Participants also valued the simplicity and effectiveness of composing text-to-image prompts. 
For example, even without prior GenAI experience, PF14 commented {\it ``it was really easy to write prompt. I love the flexibility to write prompts. It also helped me think''}.

\begin{figure*}
    \centering
    \includegraphics[width=\textwidth]{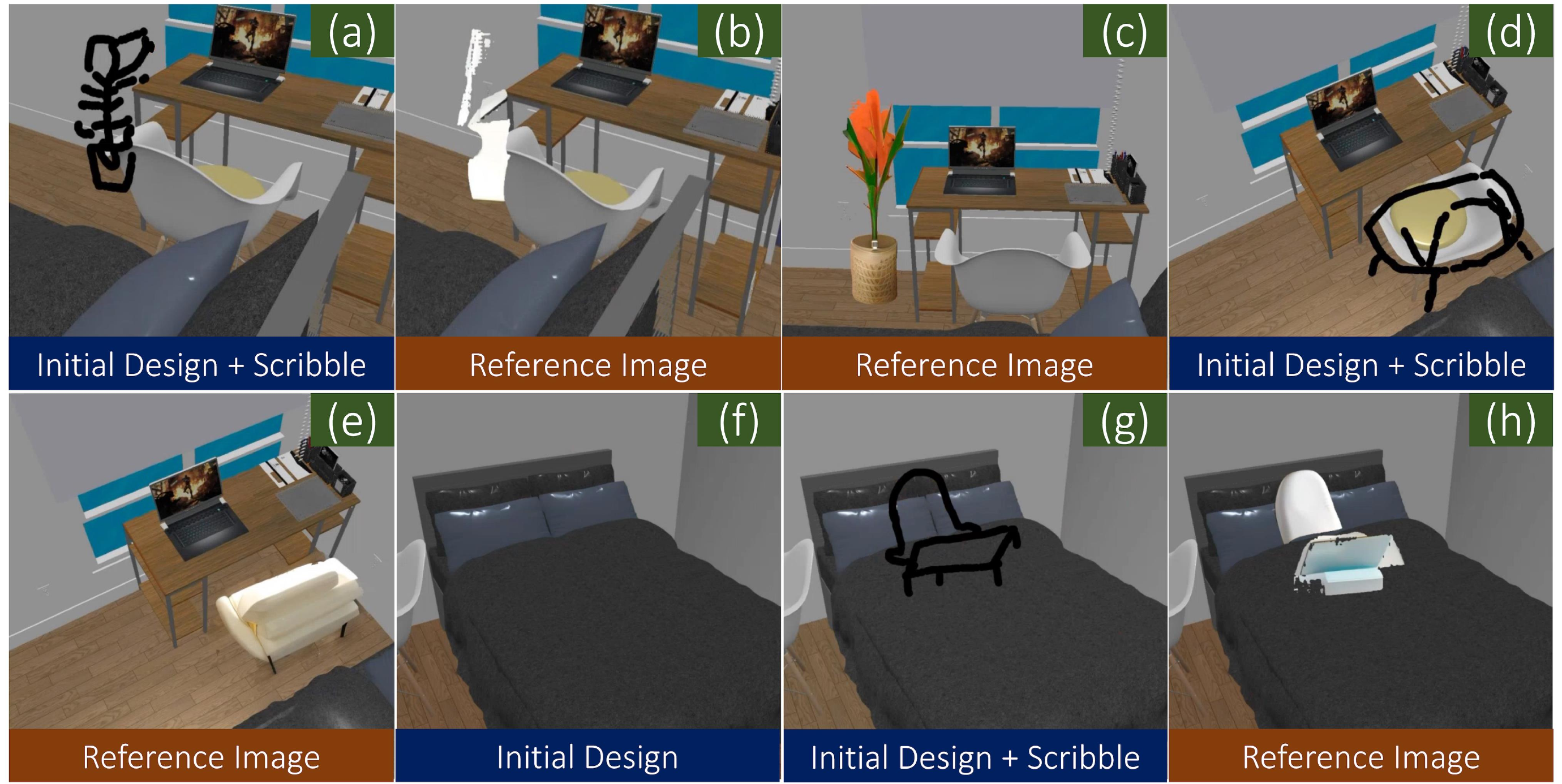}
    \caption{Examples of unsatisfactory reference images. (a) initial sketch from PF4 to describe an indoor plant. (b) \sysname~ failed to generate the desirable image. (c) with a different camera view, PF4 successfully created the expected reference image. (d) initial sketch from PF2 to describe a new chair in the office. (e) \sysname~ was able to generate a new chair but it didn't match PF2's expectation. (f - h) initial design, scribble, and final created reference image by PF7.}
    \label{fig::study-examples-failed}
\end{figure*}

\vspace{4px}\noindent\textbf{\textit{Scenarios when feedback providers fail to create explicit and satisfying companion reference images. }}
We observed multiple scenarios where participants failed to create satisfying visual references that can explicitly convey the textual feedback. 
Our analysis unveiled three key reasons.

\vspace{4px}
\noindent{\textbf{{(1)~Unsatisfactory generation.}}}
Text-to-image generation is still an emerging technology and therefore is not perfect~\cite{Liu2022}. \sysname~ could fail completely to generate a reasonable shape (Fig.~\ref{fig::study-examples-failed}b) or was able to generate a reasonable image but failed to match what the users wanted (Fig.~\ref{fig::study-examples-failed}d - e).
These setbacks may result in less explicit reference images, potentially causing misunderstandings for the designers and contributing to the negative ratings shown in Fig.~\ref{fig::study-1-results}a.
For example, PF4 thought aloud while examining Fig.~\ref{fig::study-examples-failed}b: {\it ``it doesn't look like a plant. I don't think this is explicit enough for the designer to understand''}.
In these cases, we observed that participants tend to try for several attempts or with a different view angle until they can generate the desirable image (Fig.~\ref{fig::study-examples-failed}c).
While PF4 positively rated the image explicitness, she {\it disagreed} that it was easy to create reference images.
PF2 made a similar comment with respect to Fig.~\ref{fig::study-examples-failed}e: {\it ``the chair does not look like the one I wanted''}.

\vspace{4px}
\noindent{\textbf{(2)~ Difficulty of writing prompts}.}
Few participants emphasized the importance of prompts and the challenges in writing them. 
For example, {\it ``sometimes, I need several times of revision on the prompt to make the generated image better. [...] So while my feedback takeaways could be visualized, it still needs several attempts''}~(PF12).
A small number of participants also described the needs for extracting the knowledge of the existing design: {\it ``if I make a prompt, it should have the knowledge of this existing design. For example, if I say like, keep the same blanket, then it should be same''}~(PF5).
PF11 suggested that \sysname~ should automatically create prompts based on the textual feedback: {\it ``I thought when I write the feedback, the prompt will be automatically generated so that it can bring me the reference image. But it's more like I write the feedback and then I also create the prompt [...] It was initially hard for me to distinguish the prompt and the feedback itself, that I need to tell the AI versus also tell the designer''}.
PF7, a non-native English speaker, found it challenging to phrase the prompt of ``lap desk'', leading to the complete failure of the final reference image created by \sysname~ (Fig.~\ref{fig::study-examples-failed}f - h).
This led him to rate the implicitness of the reference images (Q3) as \emph{strongly disagree}.

\vspace{+4px}
\noindent{\textbf{(3)~Lack of alternative design exploration}.}
While most participants believed that the images created by \sysname~are more explicit and could provide inspirational support compared to the baseline, some participants commented on the necessity of being able to see multiple inference results similar to mainstream search engines.
For example: {\it ``compared to the Google image, I feel there's something very inspiring about seeing like 20 images all at once from like different creators''}~(PF6),
and {\it ``when you search for an image on Google, it has the big long list of different images. That helps me to see different ideas. And because it's Google, it's pulling from a bunch of different websites. So I think that also helps me to think like, oh, this is what someone else thought. So yeah, I think if I'm thinking about it that way''}~(PF4).

\subsection{Study 2: Assessing Reference Images}\label{sec::eval::eval_feedback}
Our second study aims to tackle the key RQ: {\it how the reference images created by \sysname~could convey the gist of the 3D design feedback, compared to the images created by the baseline condition (\ie~internet searched images and/or hand sketches)?}

\vspace{+4px}\noindent{\bf Participants.}
We recruited PD1 - PD8 (age, $M = 23.75$, $SD = 2.55$, \incl~four males and four females) as the designer participants, with prior 3D design experience, from an institutional 3D design \& e\textbf{X}tended \textbf{R}eality~(XR) student society.
No designer participants were involved in the formative studies and the first study.
Details of the demographic backgrounds of designer participants can be referred to Appendix~\ref{app::evaluation_participant}.

\vspace{+4px}\noindent{\bf Procedures.}
The second study was structured as a {\it survey study}, where participants were invited to complete an online questionnaire.
We first collected all reference images collected from first study, including $44$ and $39$~design feedback, created by C1 and C2, respectively.
Each design feedback contains a textual comment, and one (or multiple) reference image(s).
We also captured the viewpoints of the initially designed 3D models for each reference image.
Next, we randomly shuffled all design feedback, which were then divided into eight groups.
All groups, except for the last one containing six design feedback, consist of $11$ design feedback each.
On average, each group comprises $46.21\%$ ($SD = 9.54\%$) design feedback created by \sysname, and $74.05\%$ ($SD = 9.21\%$) design feedback contributed by different feedback provider participants from the Study 1.
We then assigned the eight surveys to PD1 to PD8 for evaluation.
Participants were invited to rate each reference image in a $5$-point Likert scale, regarding {\it how well the reference image can convey the gist of the textual comment}.
Participants were encouraged to put their justifications as the textual comments for each rating.
Each questionnaire takes approximately $10$ - $15$~min to complete.

\vspace{+4px}\noindent{\bf Analysis.}
We analyzed the Likert-scale score for each reference image rated by one designer participants.
Qualitatively, we also collected the textual comments that participants provided to justify their ratings.
These include $33$~and $31$~textual comments for design feedback created by baseline and \sysname, respectively.
Inductive coding approach~\cite{Bingham2021} were used to analyze participants' qualitative responses.

\vspace{+4px}\noindent{\bf Results and Discussions}

\begin{figure*}[t]
    \centering
    \includegraphics[width=\textwidth]{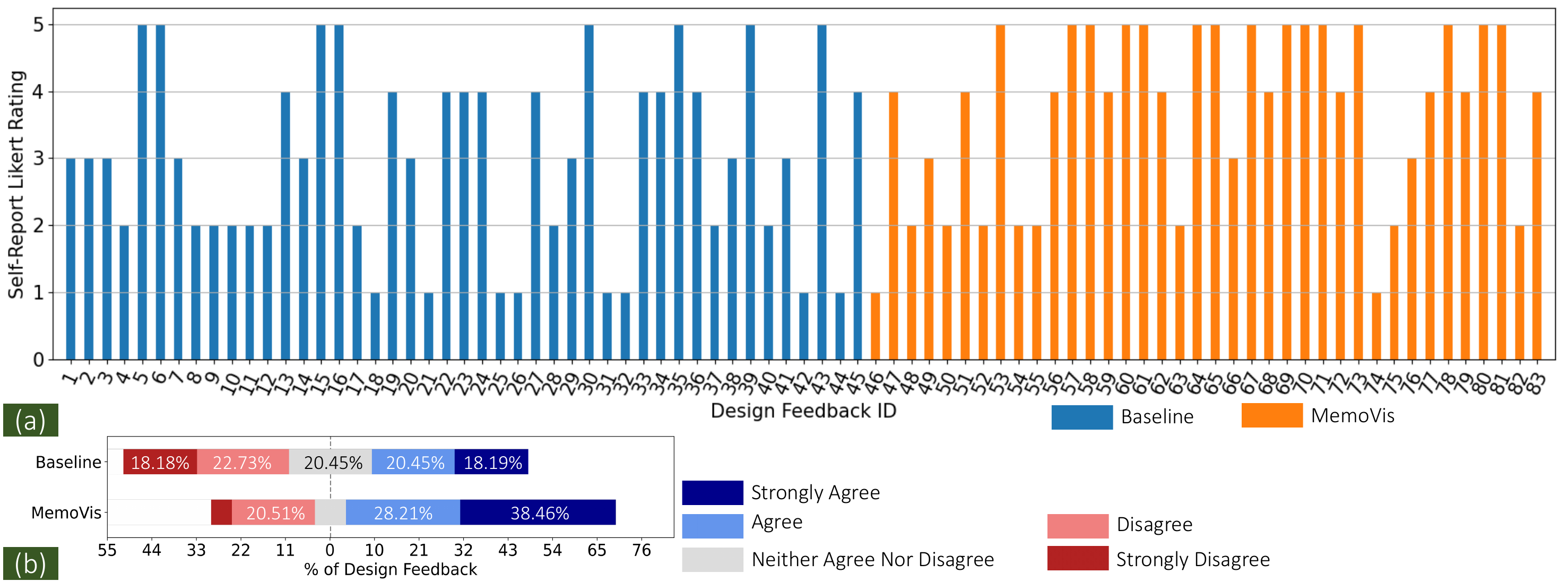}
    \caption{Survey responses of Study 2. (a) Participants' response of each 3D design feedback, in a scale of \bm{$1$} to \bm{$5$} where \bm{$1$} indicates \emph{strong disagree} and \bm{$5$} indicates \emph{strongly agree}; (b) cumulative analysis of the \% of 3D design feedback for each Likert scale by two interface conditions. Note that Fig.~\ref{fig::study-2-results}b was generated based on the survey responses shown in Fig.~\ref{fig::study-2-results}a.}
    \label{fig::study-2-results}
\end{figure*}

\noindent Fig.~\ref{fig::study-2-results}a shows the Likert rating for each feedback.
We found that around $66.67\%$ of the reference images created by \sysname ~were rated positively, versus only around $38.64\%$ of images created by baseline approach were rated positively (Fig.~\ref{fig::study-2-results}b). 

\begin{figure*}[t]
    \centering
    \includegraphics[width=\textwidth]{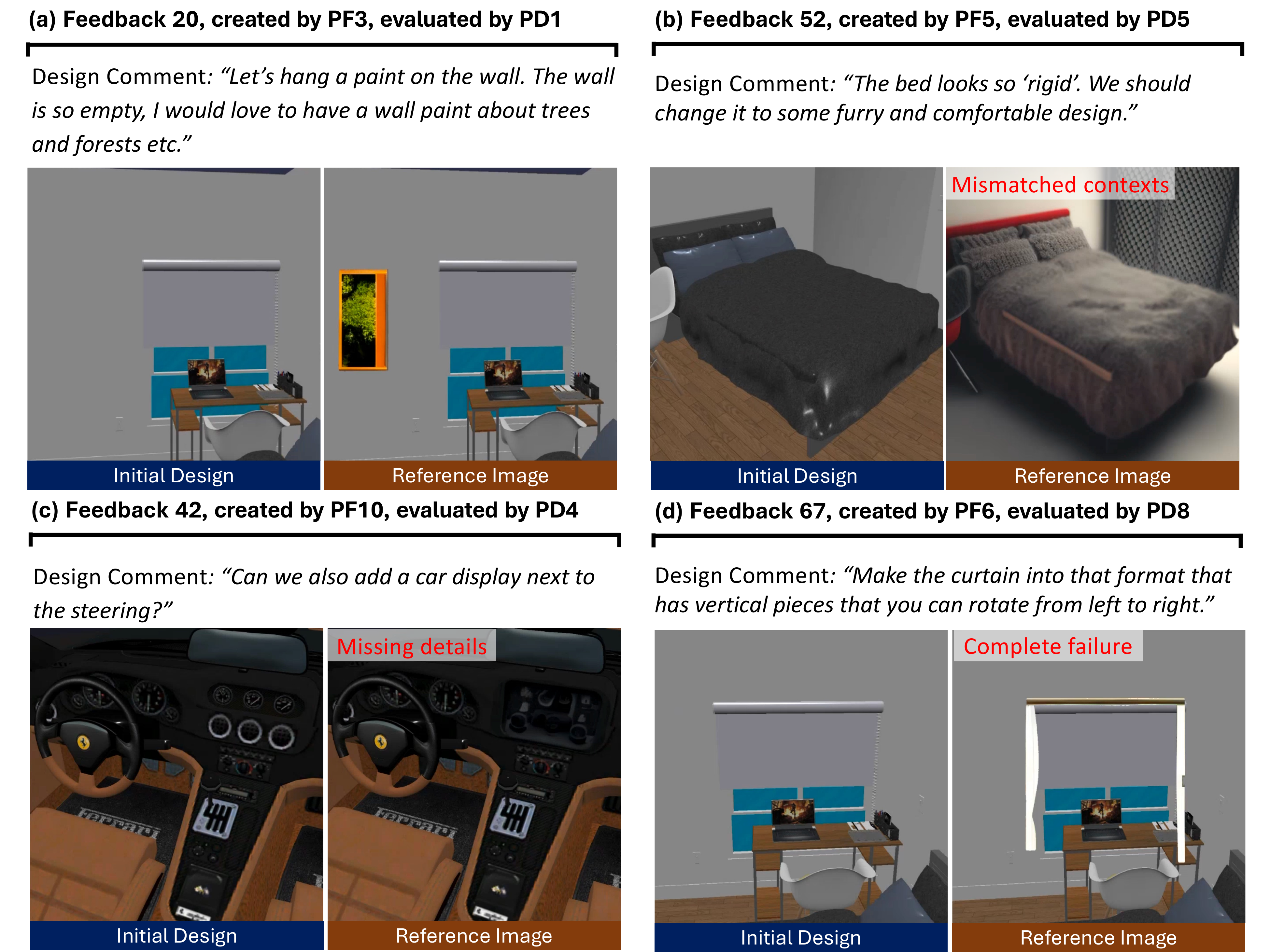}
    \caption{Examples of 3D design feedback created by \sysname. The participants \emph{strongly agreed}, \emph{neither agree nor disagree}, \emph{disagreed} and \emph{strongly disagreed} that the reference image from the 3D design feedback (a), (b), (c) and (d) can convey the gist of the textual comment, respectively.}
    \label{fig::study-2-baseline-example}
\end{figure*}

When evaluating reference images and feedback produced using \sysname ~, participants liked the explicitness of the generated reference images, \eg~ {\it ``clear and focused''}~(PD6), {\it ``the reference image perfectly captures all the elements mentioned in the design comments''}~(PD4) and {\it ``the image clearly shows what the text is trying to say''}~(PD1). 
Fig.~\ref{fig::study-2-baseline-example}a shows an example of how PD1 believed that {\it ``it is easily to identify the new picture and the desired location''} in the initial bedroom 3D model.

\begin{figure*}
    \centering
    \includegraphics[width=\textwidth]{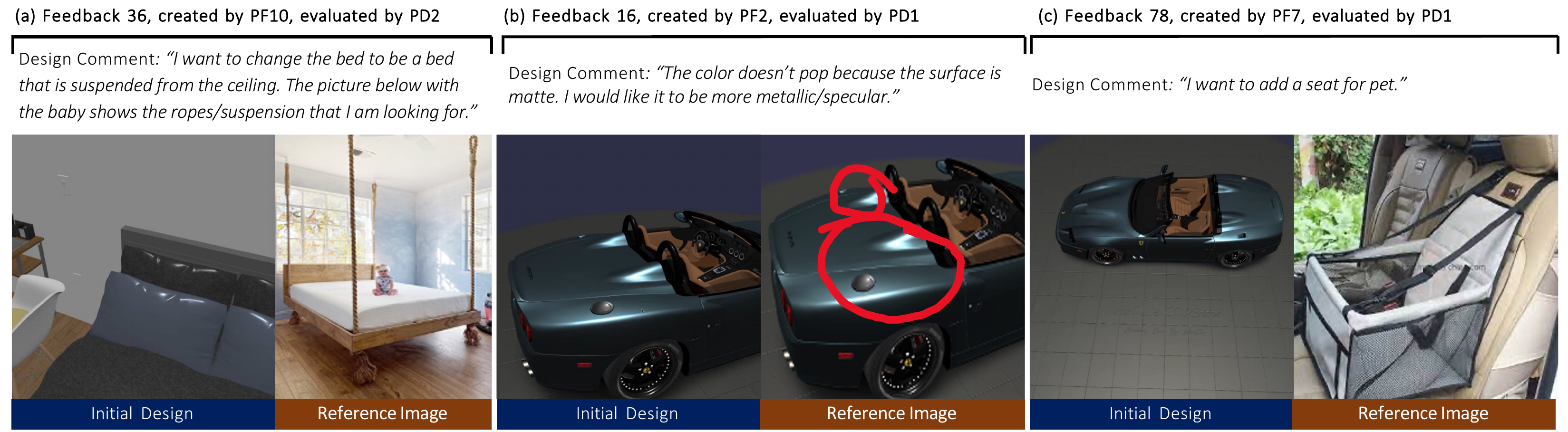}
    \caption{Examples of 3D design feedback created by the baseline interfaces. The red traces indicate the annotations drawn by the feedback provider participant PF2.}
    \label{fig::study-2-baseline-example-compare}
\end{figure*}

For the images produced in the baseline condition, comments from the designer participants point to a common drawback of mismatched contexts: the contextual difference between the reference images and the textual comments can cause confusion. 
For example, Fig.~\ref{fig::study-2-baseline-example-compare}a shows an example of the reference image created by PF10, where PD2 judged: {\it ``the structure between the bed and the floor can easily be mistaken for legs upon a cursory glance''}. 
Fig.~\ref{fig::study-2-baseline-example-compare}c shows an example when both the viewpoint and the requested change in the reference image are dramatically different to the initial design. The designer participant did not feel fully confident in grasping the feedback: {\it ``the idea of a pet seat is clear, but it seems so different from the original image that it would be confusing''}~(PD1).
We also found that some feedback provided attempted to reduce ambiguity by annotating on the initial design to highlight the changes (Fig.~\ref{fig::study-2-baseline-example-compare}b). This approach, however, is not as effective when the requested change is not explicit. In this case, the feedback provider was asking for a new material design on the car body. PD1 commented: {\it ``I'm not sure what the circles are trying to show''}.

Although most of participants favored the reference images created by \sysname, participants also highlighted few setbacks of the \sysname-generated images. 
We identified three key reasons from the designers' perspective.

\vspace{+4px}
\noindent{{\bf (1) ~Mismatched contexts}}.
Similar to the baseline condition, occasionally, designer participants pointed out context mismatches, although these did not affect their understanding of the design feedback.
Fig.~\ref{fig::study-2-baseline-example}b shows an example of how the overall bedroom design was changed though the focus of the textual feedback is the bed, which might be caused by the failure of the grab'n go modifier.
Despite this, PD5 noted: {\it ``even the background is a little different, the reference image still preserves the angle of the view, and the new environment setting''}.

\vspace{+4px}
\noindent{{\bf (2) ~Missing details}}.
While some reference images can reflect the suggested edits, designer participants believed that the missing details might cause misunderstanding.
For example, regarding Fig.~\ref{fig::study-2-baseline-example}c, PD4 \emph{disagreed} that the reference image can convey the textual comment because {\it ``while the display has been added as per the design comments, some panels have been removed. I don't know if these removals align with the design comments''}.
Such confusion might result in incorrect translations of the design feedback into the final 3D model.
While mismatched contexts were identified as common problems, designer participants did not highlight missing details for the reference images created by baselines.

\vspace{4px}
\noindent{{\bf (3) ~Complete failure}}.
In very few cases, designer participants highlighted that the reference image can be entirely unsuccessful. For example, PD8 commented on the design feedback of Fig.~\ref{fig::study-2-baseline-example}d: {\it ``I don't understand what is that white thing, doesn't look like a curtain''}.
Such reference images might cause designers to misunderstand the gist of the design feedback, potentially necessitating further communication with the feedback providers.

%% file: 6-discussions.tex
\section{Discussions}\label{sec::discussion}
Having demonstrated the \sysname~ system as an effective GenAI-powered tool for creating companion reference images for 3D design feedback, this section discusses the practical implications~(Sec.~\ref{sec::discuss::app}) and future improvements~(Sec.~\ref{sec::discuss::genai}) informed by our explorations.

\subsection{Practical Implications}\label{sec::discuss::app}

Overall, our studies showed that \sysname~ can assist feedback providers in efficiently creating companion reference images for 3D design feedback.
The design of \sysname ~ validates the feasibility of using GenAI and VLFMs to support a more simple and efficient workflow of asynchronous 3D design review. 
This section discusses key practical implications drawn from our findings.

\vspace{+4px}\noindent{\bf Real-time viewpoint suggestions}.
To assist in creating reference images, \sysname ~ needs to first allow feedback providers to efficiently locate 3D camera viewpoints pertinent to the written textual comments.
Our study showed that the real-time viewpoint suggestions can support this task by analyzing the written comment and suggesting semantically-relevant views in the 3D scene. 
The effectiveness of this feature has also been demonstrated by $11$ feedback providers {\it without} proficient 3D software skills.
Despite having 3D skills, few participants like PF2 found this feature valuable when it comes to seeking relevant views on a potentially large-scale 3D model.
While earlier studies suggested closed-form solutions for viewpoint selection based on area~\cite{Plemenos1996}, silhouette~\cite{Feldman2005, Vieira2009}, and depth~\cite{Blanz1999} attributes, along with their combinations~\cite{Secord2011}, the relationships between the chosen view and textual semantics are often absent.
\sysname~ introduces a novel paradigm, enabling feedback providers to effortlessly identify relevant views for contextualizing the reference images they intend to create.
Further, looking at the qualitative feedback from Sec.~\ref{sec::eval::create_feedback}, a number of participants, exemplified by PF2, believed that viewpoint suggestion would be even more useful in larger-scale 3D models such as those used in video game design.
While prior research, such as IsoCam~\cite{Marton2014}, has explored the viability of employing a touch-based controller for navigating the camera in large projection setups, it is impractical to apply such complex hardware setups to 3D design review workflow. 
Instead, \sysname~ leverages VLFMs to infer what types of views that the feedback providers might be interested in exploring, enabling feedback providers to spend less time on maneuvering the viewing camera and more time on the main {\it feedback writing} task.

\vspace{+4px}
\noindent{\bf Image modifiers.}
We highlighted the advantages of \sysname's rapid image creation workflow with three types of image modifiers, for facilitating creating companion reference images while writing 3D design feedback.
It is worthwhile to emphasize that \sysname ~ is a review and feedback creation tool, instead of an image editing tool like the recent GenAI-powered Photoshop~\cite{Photoshop, photoshopAI}, or an ideation tool like Vizcom~\cite{Vizcom}.
Hence, advancing techniques to realize aesthetic and high-quality images are \emph{beyond} our scope. Rather, our key focus is to prioritize the clarity of the reference images and their alignment with textual comments.
Our results have shown the explicitness of the reference images created and the capabilities to maintain the contexts of the anchored viewpoint, confirmed by both feedback providers (Sec.~\ref{sec::eval::create_feedback}) and designers (Sec.~\ref{sec::eval::eval_feedback}).
Additionally, \sysname's approach can also  help with creative thinking. The images generated with the modifiers can sometime offer new ideas and inspirations for feedback writers.
We also demonstrated that a larger number of participants found the reference image creation workflow using \sysname ~to be easier compared to today's methods involving image searching and/or sketching (Fig.~\ref{fig::study-1-results}a).
While the design of \sysname ~is intended to encourage feedback providers to focus on the feedback typing task, we showed that it is still viable to request feedback providers to engage in simple rough sketching and painting directly on top of the 3D explorer.
Our discovery indicates that incorporating simple non-textual input modalities like sketches and painting will not significantly raise the interaction cost~\cite{Raluca2013}. Instead, it gives feedback providers added control to ensure consistency in the reference image within the context of the initial design, as highlighted in our formative studies.
In theory, this finding could be linked to Kohler's recommendations~\cite{Kohler2022} for generic user experience design: granting users autonomy through customization may enhance the sense of ownership, potentially improving the overall interaction experience. Nevertheless, dedicating time and resources to customization development could also elevate interaction costs, possibly diminishing the user experience~\cite{Raluca2013, Heidi2008, Norman2013, Bennett2023}. 
\sysname~ presents an exemplary design that balances low interaction costs while affording feedback providers additional controls in the creation of reference images.

\vspace{+4px}
\noindent{\bf Integration with the State-Of-The-Art (SOTA) models.}
We demonstrated the feasibility of using the \sysname ~ system, powered by year-2023's pre-trained models, to assist feedback providers in efficiently creating reference image for 3D design feedback.
With continuous advancement of SOTA performance of today's VLFMs~\cite{Morris2023AI, Chen2023AI}, we believe our contribution toward designing a novel interaction workflow and experience for efficient 3D design review will continue to hold.
We consider the underlying GenAI and VLFMs as the engineering primitives to drive the novel interaction experience, where the feedback providers could efficiently create companion reference images for the feedback comments, while focusing on text typing.
The enhancements of the inference quality of recent text-to-image SOTA models could help generalizing the practical applicability of \sysname ~ through potentially more photorealistic synthesized images ~\cite{Balaji2023}, simpler and more intuitive prompts ~\cite{Hao2022}, as well as a reduced inference latency ~\cite{Yang2023}.
For example, SOTA pipelines, such as Promptist~\cite{Hao2022} for optimizing text-to-image GenAI prompts, could potentially reduce the failures when feedback providers write low-quality prompts.
Other 3D-related GenAI SOTA pipelines like InseRF~\cite{InseRF} that enables text-driven 3D object insertions might also be integrated to enhance features for feedback providers \textit{with} prior 3D experience, enabling deeper exploration.

\vspace{+4px}
\noindent{\bf Integration with real-world workplace applications.} 
The interaction designs of \sysname ~ can be integrated into larger workplace applications, as a form of lightweight plugins.
Feedback providers could focus on their primary task, {\it feedback typing}, instead of editing images or 3D models (Sec.~\ref{sec::system}).
These \sysname-translated plugins enable feedback providers to inspect 3D models, create and visualize textual feedback comments, and send them to designers just like the memo notes. 
Such interaction experience would be simple and fluid~\cite{Elmqvist2011} that facilitates smooth discussion and design collaborations between feedback providers and designers, by encouraging the feedback providers focusing on thinking and typing textual comments.
For example, \sysname~can be integrated as an add-on for Gmail.
This integration allows feedback providers to effortlessly create accompanying reference images while typing textual comments within an email.
\sysname ~ can also be implemented as a plugin for iMessage.
Feedback providers could use this plugin to conveniently examine 3D models and create reference images for textual comments while  engaging in text conversations with their designers, making the process as straightforward as creating a memoji~\cite{memoji}.
Although \sysname ~ was contextualized in an {\it asynchronous} 3D design review workflow, this plugin is invaluable in expediting the creation of reference images during {\it synchronous} conversations in \textbf{I}nstant \textbf{M}essaging~(IM) applications, effectively minimizing the duration of silence~\cite{Li2023, Kim2017}.
Beyond supporting 3D design feedback, \sysname~ can still be generalized to broader multi-modal GenAI-based applications that requires efficient visualizations of texts. For example, \sysname ~ can be integrated with existing collaborative writing tools like Notion. \sysname's capabilities of efficiently visualizing textual content shows promise in enhancing synchronous discussions without disrupting the flow of conversation.

\subsection{Improving \sysname}\label{sec::discuss::genai}

After exploring the practical implications of \sysname, this section explores potential key directions for future improvements, drawing insights from our findings.

\vspace{+4px}
\noindent{\bf Boosting AI inferences with human feedback.}
In Study 1, we observed that the feedback providers might need to adjust the viewpoint based on \sysname's suggestion and/or make multiple attempts to create the reference images that could convey the gist of the design feedback. 
One future direction is to understand how we could leverage the behavioral actions from feedback providers to boost future AI inferences.
Similar ideas have been successfully used in many large language model applications through designing prompts with few-shot learning~\cite{fewshot} and integrating the strategies of reinforcement learning from human feedback~\cite{Ziegler2020}.
For example, instead of searching possible viewpoints solely based on pre-trained CLIP model~\cite{CLIP}, \sysname ~ might incorporate the {\it past view preferences} from the feedback providers to find the most likely viewpoint, with which the feedback providers want to anchor the textual comments.
Realizing this may require the design of an effective cost function with respect to CLIP inferences and the preferences from feedback providers that the \sysname ~ could optimize.

\vspace{+4px}
\noindent{\bf Eliminating manual prompt writing toward simple and fluid reference images creation experience.}
\sysname ~ necessitates feedback providers to create additional prompts for creating reference images, which can be redundant.
Sometimes, prompts may require feedback providers to include additional context beyond the immediate objects of focus. 
Although most participants were satisfied with the experience of using \sysname~to create companion reference images for textual feedback, we found participants occasionally found it difficult (PF5) and tedious (PF12) to write additional prompt, or confused of the difference between feedback comments and prompts (PF11).  
While prior research, \eg~\cite{Hao2023, Manas2024}, has demonstrated novel prompt optimization algorithms for vanilla text-to-image GenAI, we opted not to incorporate this feature into the current implementation due to lack of validations on creating photorealistic images using conditioned text-to-image GenAI (Sec.~\ref{sec::related::genai}).
Additionally, granting flexibility to feedback providers to write prompts also enables them to iteratively enhance the prompts upon the less optimal reference images.
Although Liu~\etal~\cite{Liu2022} have discussed the key guidelines of crafting text-to-image prompt, future work might explore how to optimize the text-to-image prompts for conditioned text-to-image GenAI, and how to integrate broader contexts from the written feedback, without laborious trial-and-error process for creating prompts, as highlighted by PF10.
While advancing techniques for prompt creations and explorations is beyond our scope, future work may explore the feasibility of integrating interactive prompt engineering techniques~\cite{Feng2024, Wang2024} to assist feedback providers in streamlining the feedback creation process while allowing the exploration of the nuances of prompt creation.
Ultimately, we envision \sysname ~ being able to automatically generate the prompts for text-to-image GenAI without the awareness of feedback providers, leading to a simple and fluid interactions experience~\cite{Elmqvist2011}, where the candidate reference images could be responsively created and updated as the feedback providers typing the textual comments.

\vspace{+4px}
\noindent{\bf Integrating GenAI with searching.}
While our research on \sysname~ shows how \mbox{text-to-image} GenAI can be a potential path to enhance the 3D design review workflow, the Study 1 indicated that GenAI alone might be insufficient.
Despite most participants are satisfied with the experience of using \sysname~to create reference image for textual comments, few participants (\eg~PF6) emphasized the opportunity to {\it integrate} rather than to {\it replace} internet search and hand annotation approaches with \sysname.
While \sysname ~ is helpful when feedback providers do not have a specific design in mind, few participants (\eg~PF4, PF11) emphasized the usefulness of using online images when a specific design suggestion in the mind.
With these observations, a compelling research path naturally emerges: how could we woven today's image searching approach into \sysname's pipeline. 
Although this direction is similar to recent works like GenQuery~\cite{Son2023GenQuery}, which showed how to integrate GenAI and search to help instantiate designers' early-stage abstract idea, and DesignAID~\cite{Cai2023}, which emphasized the importance of {``augmenting humans rather than replacing them''} while attempting to use GenAI to help exploring visual design spaces, reviewing and providing feedback for 3D design are fundamentally different from 2D graphic ideation workflow due to the complexities of 3D models and the convoluted thinking process while feedback providers are attempting to create reference images (Sec.~\ref{sec::eval::create_feedback}).
Future direction might consider how to integrate existing search engine into \sysname, opening new paradigms of efficiently co-visualizing textual feedback alongside human interactions, internet searches and GenAI.
For example, despite the imperfections and the potential risk of causing misunderstandings, our results indicate that a few feedback providers still prefer to use online images tailored to their specific design needs.
While tools like Photoshop~\cite{Photoshop, photoshopAI} enable feedback providers to edit online images, such workflows are often not streamlined, time-consuming and inefficient for those lacking image editing skills.
Although \sysname ~ empowers feedback providers to use image modifiers to create reference images primarily from text, future work may further explore how these image modifiers can be extended and integrate the contexts of the searched image(s).

\section{Limitations}\label{sec::limitation}
Having demonstrated the promising of \sysname, we also acknowledge multiple key limitations with respect to system design and evaluations.

\vspace{+4px}
\noindent\textbf{Inference latency.}
\sysname~ took around $30$~seconds before generating a synthesized image.
While \sysname~is asynchronous, through which the feedback providers could continuous explore the 3D model, we found most participants still prefer to shorten the wait time for a more streamlined feedback creation workflow.
Although reducing latency is beyond our scope, some participants (\eg~PF6) found it as a critical setbacks in terms of software engineering design perspective, compared to the internet searching.
We speculate future advancement of SOTA GenAI and GPU parallel computing research would help overcome this limitation.

\vspace{+4px}\noindent\textbf{Participants.}
Our first formative study was conducted with only two participants as it required individuals with extensive design experience.
The evaluation studies were conducted with only $14$ feedback providers and eight designer participants, due to the lengthy duration in Study 1 and limited resources for Study 2 which requires participants to have prior 3D design experience.
Despite the validity of our results that were mainly qualitative based, future work might further explore the usability of \sysname~ with more participants coming from different backgrounds, to minimize the bias from the recruited participants.
For instance, in our study tasks, the feedback provider participants recruited can only represent the clients engaged with the workflow of design feedback creations. Future research may recruit feedback provider participants who can represent other types of feedback providers like collaborators and managers.

\vspace{+4px}\noindent\textbf{Evaluation conditions and tasks.}
First, our studies were only based on a bedroom and a car model (Appendix~\ref{app::study_tasks}).
Future researchers might explore the generalizability of our system to a wider ranges of 3D models.
As discussed in Sec.~\ref{sec::discuss::app} and speculated by PF2 that the viewpoint suggestions could be {\it ``much helpful''} for a larger scene, future work might also investigate how \sysname~ could help feedback providers to navigate and create reference images for a larger environment 3D models.
Second, the baseline condition in the Study 1 (Sec.~\ref{sec::eval::create_feedback}) required participants to create reference images using searching and/or hand sketching.
Although, through our formative studies, it is possible to simulate the prevailing practices of most participants in creating reference images for 3D design feedback, future research might explore comparing \sysname ~ with a broader range of GenAI-powered baseline, such as creating reference images using vanilla text-to-image GenAI tools~\cite{AdobeFirefly} and professional image editing software~\cite{Photoshop}.

\vspace{+4px}\noindent\textbf{Evaluations in an ecologically valid 3D design review workflow.}
Despite the potentials of integrating \sysname ~ as part of larger collaborative design workflow and workplace applications (Sec.~\ref{sec::discuss::app}), our current evaluation was based on a monolithic browser-based application in a controlled laboratory setting. 
Future research might deploy \sysname, and investigate the user experience in a realistic real-world 3D design review workflow.
One future direction might investigate the feasibility and effectiveness of \sysname, after being integrated and deployed into today's mainstream workplace and IM applications such as Gmail and iMessage~(Sec.~\ref{sec::discuss::app}).

%% file: 7-conclusion.tex
\section{Conclusion}\label{sec::conclusion}
We designed and evaluated \sysname, a browser-based text editor interface that assists feedback providers in easily creating companion reference images for textual 3D design comments. \sysname~ integrates several AI tools to enable a novel 3D review workflow where users could quickly locate relevant design context in 3D and synthesize images to illustrate their ideas.
A within-subject study with $14$~feedback providers demonstrates the effectiveness of \sysname.
The quality and explicitness of the companion images were evaluated by another eight participants with prior 3D design experience.

%% file: ack.tex
\begin{acks}
    We thank the insightful feedback from the anonymous reviewers.
    We appreciate the discussions with fellow researchers from Adobe Research, including Mira Dontcheva, Anh Truong, Joy O. Kim, and Zongze Wu, as well as Rima Cao from UC San Diego.
    We thank Zeyu Jin for providing the text-to-voice GenAI pipeline to synthesize narrations in the companion videos.
\end{acks}

%% file: A01-ethical-disclaimer.tex
\section{Ethical and Copyright Disclaimer}\label{sec::app::ethical_disclaimer}
This work has been approved by the \textbf{I}nstitutional \textbf{R}eview \textbf{B}oard (IRB). 
All \textbf{P}ersonal \textbf{I}dentifiable \textbf{I}nformation (PII) has been removed.
Prior each user study, we went through the approved informed consent with all participants and have obtained participants' consent on video, audio, and screen recordings.
While monetary incentives were not provided, all participants were given opportunity to try out and experience of SOTA GenAI technologies as the time of writing, and to know further about our research.
For research and demonstration purposes, we used multiple internet searched images in Fig.~\ref{fig::polycount-example} and Fig.~\ref{fig::study-2-baseline-example}, copyrights of these images belong to the original content creators. 
While copyrights of the synthesized images created by GenAI are still a challenging research topics~\cite{Samuelson2023, Murray2023}, we do not claim the copyrights of all synthesized images by GenAI. 
These images are only used for research and demonstration purposes.

%% file: A03-controlnet-depth-scribble.tex
\clearpage
\section{Examples of ControlNet with Depth and Scribble Conditions}\label{sec::app::compare_controlnet_depth_scribble}
\begin{figure*}[h]
    \centering
    \includegraphics[width=\textwidth]{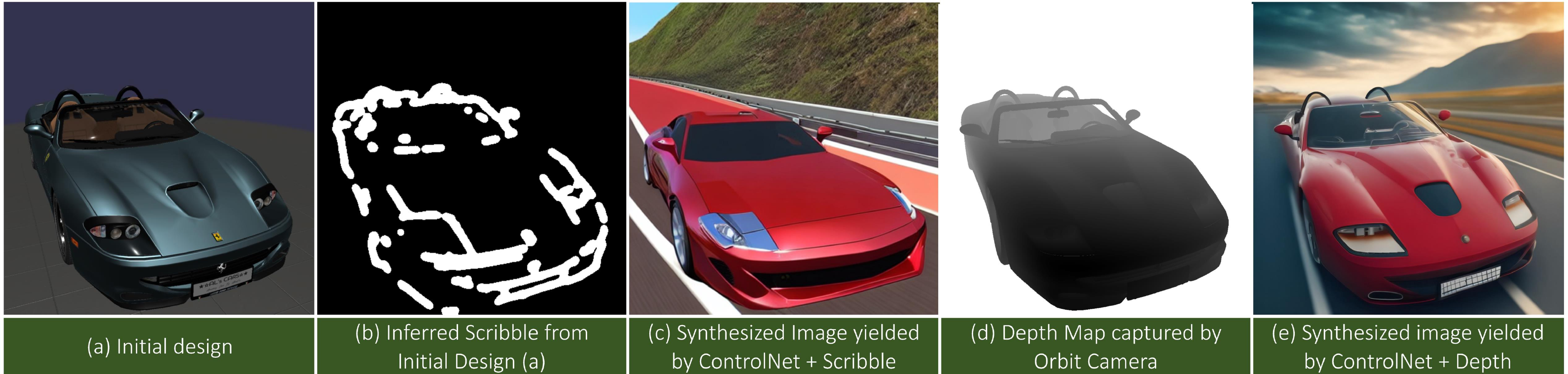}
    \caption{Examples of synthesized images created by ControlNet conditioned on scribble and depth. (a) The viewpoint of the initial 3D design; (b) Users' scribble inferred from (a); (c) Synthesized image yielded by ControlNet conditioned on scribble; (d) Depth map captured by orbit camera with (a); (e) Synthesized image yielded by ControlNet conditioned on depth. For both conditions, we used the prompt {``a red car driving on the freeway''}.}
    \label{fig::controlnet-scribble-depth-compare}
\end{figure*}

Designing of \sysname ~ leveraged the year-2023's pretrained ControlNet conditioned on depth and scribble~\cite{Zhang2023ControlNet}.
This section provides additional supplementary example demonstrating how depth-conditioned ControlNet could better anchor the generated image based on the original design.
Fig.~\ref{fig::controlnet-scribble-depth-compare}c and Fig.~\ref{fig::controlnet-scribble-depth-compare}e shows an example of how the synthesized images look like guided by the textual prompt {\it ``a red car driving on the freeway''}.
Notably, Fig.~\ref{fig::controlnet-scribble-depth-compare}b and Fig.~\ref{fig::controlnet-scribble-depth-compare}d shows the inferred scribbles with HED annotator~\cite{Xie2015HED} and the depth map of the 3D model.

%% file: A02-algorithm.tex
\section{Supplementary Algorithm Design}\label{sec::app::algo}

Algo.~\ref{alg::remove_hidden_obj} shows the algorithm to remove the residual pixels when a new objects are added by the feedback providers, with text + scribble and grab'n go modifiers.
Full system design details could be referred to Sec.~\ref{sec::system::modifiers}.

\begin{algorithm}[h!]
   \small 
   \caption{Approximate initial design image without hidden objects.} \label{alg::remove_hidden_obj}
   \begin{algorithmic}[1]

        \PROCEDURE{SelectMeshPrimitives}{$\bm{I}_{seg}$, $\bm{v}$, $r_{th} = 0.7$}
            \STATE $cam =$ \textbf{new} $OrbitCamera()$
            \STATE $cam.setViewAngle(\bm{v})$
            \STATE $scene.render()$
            \STATE $hits \gets$ \textbf{new} $Set()$
            \FOR{{sampled pixels \textbf{in}} $\bm{I}_{seg}$}
                \IF{$\bm{I}_{seg}(r, c) == 0$}
                   \STATE \textbf{continue}
                \ENDIF
                \STATE $mesh \gets scene.raycast(r, c, cam)$
                \IF{$mesh != null$}
                   \STATE $hits.add(mesh)$
                \ENDIF
            \ENDFOR
            \FOR{$mesh$ \textbf{in} $hits$}
                \STATE $DepthTextureRenderer.renderList \gets [mesh]$
                \STATE $DepthTextureRenderer.render()$
                \STATE $depth \gets DepthTextureRenderer.getImage()$
                \STATE $\bm{I}_{mesh} \gets (depth < depth.max())$
                \STATE $r \gets sum({\bm{I}_{mesh} \cap \bm{I}_{seg}}) /  sum(\bm{I}_{mesh})$
                \IF{$r \leq r_{th}$}
                   \STATE $hits.remove(mesh)$
                \ENDIF
            \ENDFOR
            \STATE \textbf{return} $hits$
        \ENDPROCEDURE

        \PROCEDURE{GetInitialImage}{$\bm{I}_{seg}$, $\bm{v}$, $r_{th} = 0.5$}
            \STATE $meshes \gets SelectMeshPrimitives(\bm{I}_{seg}, \bm{v}, r_{th} = 0.5)$
            \STATE $RGBTextureRenderer.renderList \gets meshes.toList()$
            \STATE $RGBTextureRenderer.render()$
            \STATE \textbf{return} $RGBTextureRenderer.getImage()$
        \ENDPROCEDURE
    \end{algorithmic}
\end{algorithm}

%% file: A04-pretrained-models.tex
\clearpage
\section{Pre-Trained Models}\label{sec::app::models}
\sysname~ was prototyped based on a set pre-trained model, and deployed on a cloud server with four A10G GPUs.
This section provides supplementary details of the VLFMs we used.
Full design and implementation details could be referred to Sec.~\ref{sec::system}.

\begin{itemize}[noitemsep, leftmargin=*]

\item {\bf CLIP}. We used the pre-trained \texttt{clip-ViT-B-32} due to its inference performance and our available computing resources. Other variant CLIP pre-trained model would lead to similar results.

\item {\bf SAM}. We used the checkpoint of \texttt{sam\_vit\_h\_4b8939.pth}. All hyper-parameters that were used during inference are consistent with the default suggestions~\cite{Kirillov2023SAM}.

\item {\bf ControlNet}. We used two different pre-trained ControlNet models. For depth-conditioned synthesis, we used an internal pre-trained model developed by our organization. 
Similar to the original ControlNet model, our internal model generates realistic images from the depth maps. For our use cases, we opted to use this internal model since it was optimized to generate high quality realistic images. 
While the open-source equivalent of depth-conditioned ControlNet (\texttt{lllyasviel/sd-controlnet-depth}) might work, the quality of the synthesized images might be degraded.
For depth- and scribble-conditioned synthesis, we used the open source pre-trained ControlNet under the depth (\texttt{lllyasviel/sd-controlnet-depth}) and scribble (inferred by HED annotator~\cite{Xie2015HED}) conditions, based on original Stable Diffusion (runwayml/stable-diffusion-v1-5). 
For all inference tasks, we used {\it ``realistic, high quality, high resolution, 8k, detailed''} and {\it ``monochrome, worst quality, low quality, blur''} as the positive and negative prompts, respectively. 
The inference step was set to $30$ to balance the quality of synthesized image and inference latency.

\item {\bf Inpainting}. The model \texttt{kandinsky-community/kandinsky-2-2-decoder-inpaint} was used to support inpainting feature. The prompt {\it ``background''} was used to realize the feature of removing object(s) from the reference image~\cite{inpaintRemove}.

\end{itemize}

%% file: A05-participants.tex
\clearpage
\section{Participants Recruitment for User Studies}\label{app::evaluation_participant}
\begin{figure}[h!]
    \centering
    \includegraphics[width=\textwidth]{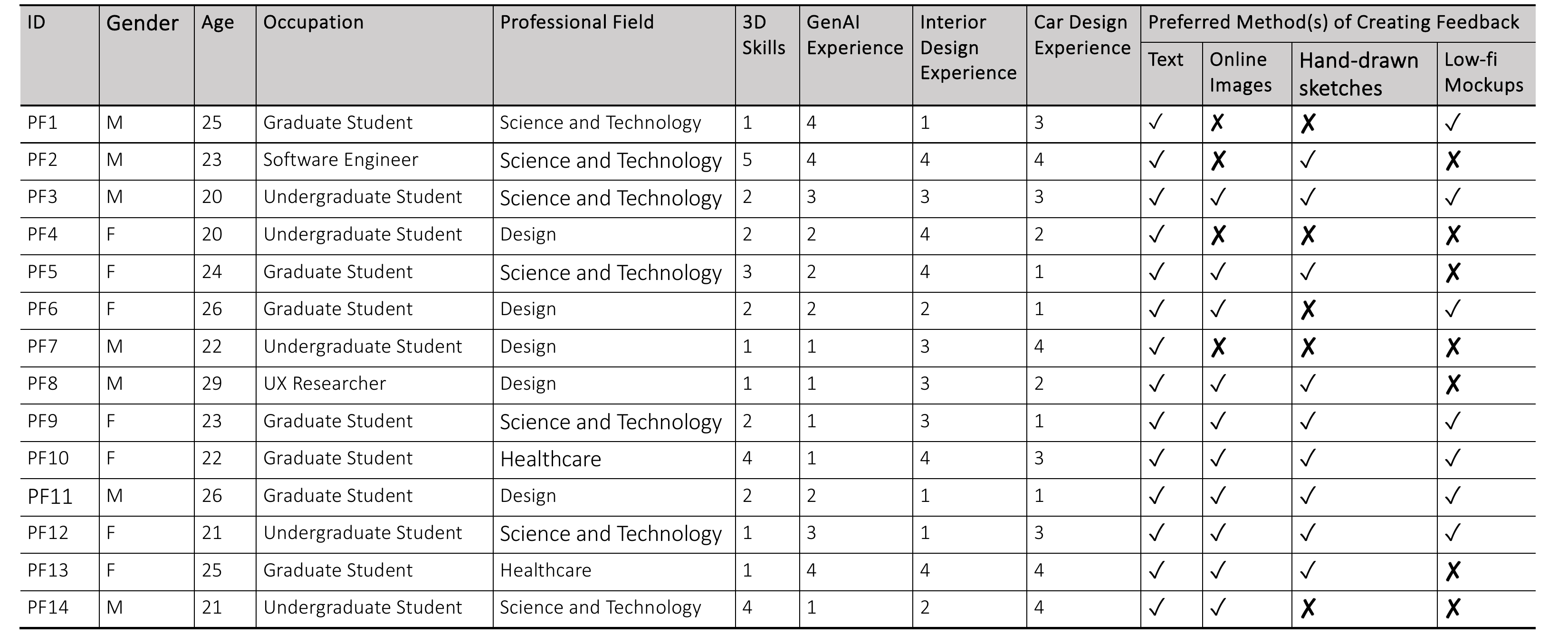}
    \caption{Participants' demographic background of the Study 1 (Sec.~\ref{sec::eval::create_feedback}). In a 5-point Likert scale, participants were instructed to self-evaluate their skills of using 3D software and prior experience of GenAI, interior design as well as car design. The methods of ``text'', ``online images'', ``hand-drawn sketches'', and ``low-fi mockups'' refer to the feedback communication method of using typed texts, online-searched reference images, hand-drawn sketches, and the reference images created using image editing tools.}
    \label{fig::study1-participant}
\end{figure}

Fig.~\ref{fig::study1-participant} shows the participants' demographic background of the Study 1 (Sec.~\ref{sec::eval::create_feedback}).
Participants were recruited from various social media platforms and a university instant messaging application group, \incl~ faculty, students and alumni, through convenience sampling.
Even though all participants viewed themselves as experienced in providing design feedback, we reported their preferred methods for creating feedback in the past.
Fig.~\ref{fig::study1-participant} shows that all participants used texts to communicate the feedback.
Participants may also use reference images searched from online, drawn by hand sketches, and/or created using image editing tools, which are referred as ``online images'', ``hand drawn sketches'', and ``low-fi mockups'' respectively in Fig.~\ref{fig::study1-participant}.
The power analysis of the sample size of $14$ demonstrates a power of $80.06\%$, with the significance level ($\alpha$), number of groups, and effect size being set to $0.05$, $2$, and $0.4$, respectively. 
We used the $\eta_p^2$ to compute the effect size, and chose a large effect size due to the lengthy duration of the Study 1.

\begin{figure}[h!]
    \centering
    \includegraphics[width=\textwidth]{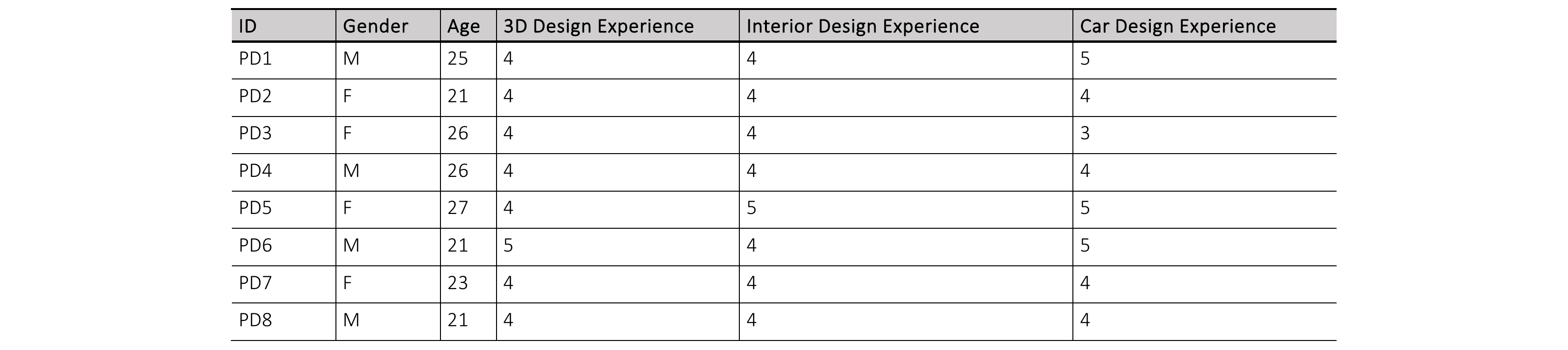}
    \caption{Participants' demographic background of the Study 2~(Sec.~\ref{sec::eval::eval_feedback}). In a scale of 1 to 5, participants were instructed to self-evaluate their prior experience of 3D design in general as well as experience of interior and car design.}
    \label{fig::study2-participant}
\end{figure}

Fig.~\ref{fig::study2-participant} shows the participants' demographic background of the Study 2 (Sec.~\ref{sec::eval::eval_feedback}).
In a scale of $1$ to $5$, participants were instructed to self-rate their prior experience of 3D design in general, interior as well as car design.

%% file: A06-study-task.tex
\clearpage
\section{Study Tasks}\label{app::study_tasks}
\begin{figure}[h]
    \centering
    \includegraphics[width=\textwidth]{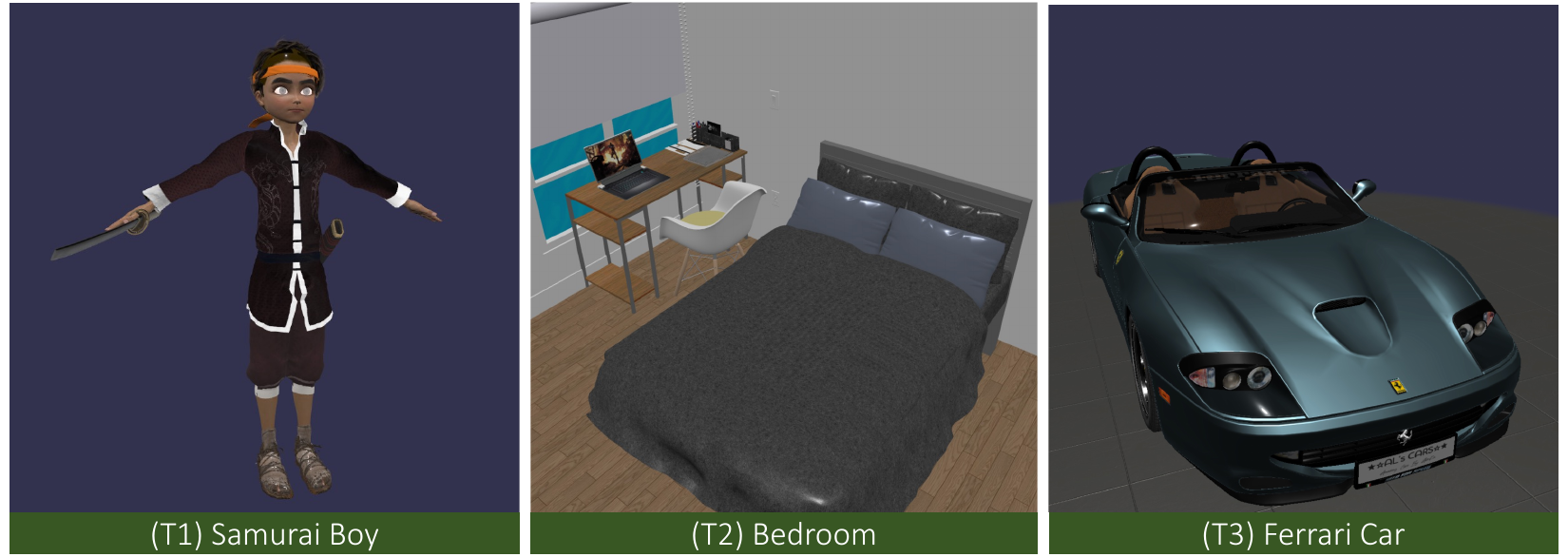}
    \vspace{-0.25in}
    \caption{3D models used in the user study, including a samurai boy (for task T1), a bedroom (for task T2), and a car (for task T3). Notably, T1 was only used for training.}
    \vspace{-0.1in}
    \label{fig::study-models}
\end{figure}

This section provides supplementary materials of the study tasks that were used in final evaluations (Sec.~\ref{sec::evaluation}).
Fig.~\ref{fig::study-models} shows three models we used for feedback providers to review and create textual feedback comments with companion visual reference image(s).
With T1, participants were required to provide feedback for a samurai boy design (Fig.~\ref{fig::study-models}a).
This tasks were used to help feedback provider participants to get familiar with both interface conditions.
All data collected from T1 was \emph{excluded} from the final analysis.
With T2, feedback provider participants were instructed to enhance the bedroom design in Fig.~\ref{fig::study-models}b, such that the bedroom is comfortable to live in.
With T3, feedback provider participants were instructed to improve the design of a car in Fig.~\ref{fig::study-models}c, such that the car is comfortable to drive with (in terms of aesthetics, ergonomics, and functionalities).
Full study details could be referred to Sec.~\ref{sec::evaluation}.

%% file: A07-modifier-usages.tex
\section{Usages of the Image Modifiers}\label{sec::app::modifier_usages}
\begin{figure}[h]
    \centering
    \includegraphics[width=\textwidth]{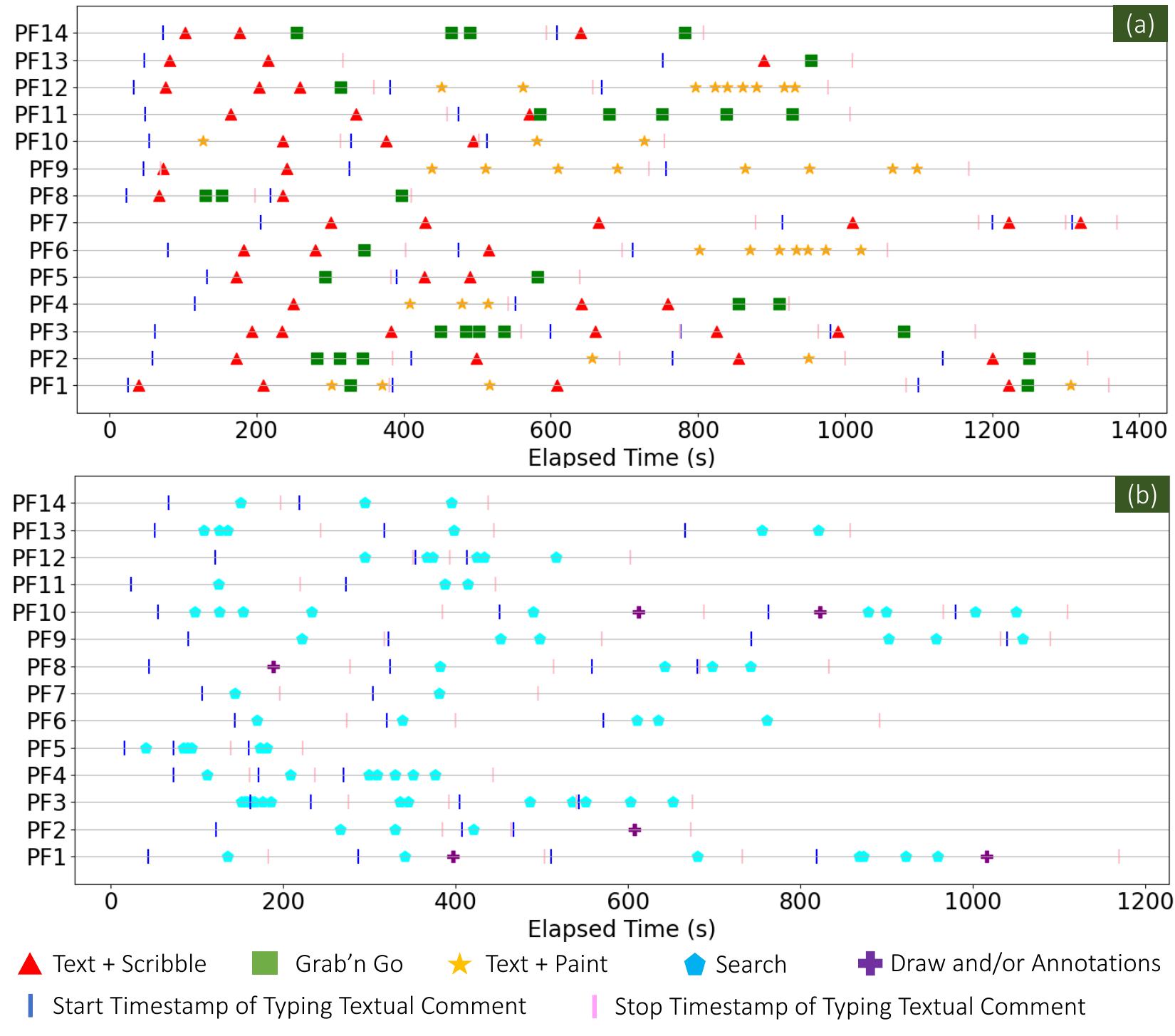}
    \caption{Visualizations of how each participants interact with one (or multiple) image modifiers, with \sysname ~(a) and baseline~ (b) interface conditions.}
    \label{fig::study-1-modifier-usages}
\end{figure}

As a supplementary material for Sec.~\ref{sec::eval::create_feedback}, Fig.~\ref{fig::study-1-modifier-usages} shows the visualizations of how each participants used the one (or multiple) of the image modifiers, provided by \sysname.
For the baseline condition, we visualize the instants when the feedback providers use search, as well as draw and/or annotations, to create reference images.

%% file: A08-codebook-online-analysis.tex
\section{Codebook and Themes from Qualitative Data Analysis}\label{sec::app::codebook}
\begin{figure*}[h]
    \centering
    \includegraphics[width=0.65\textwidth]{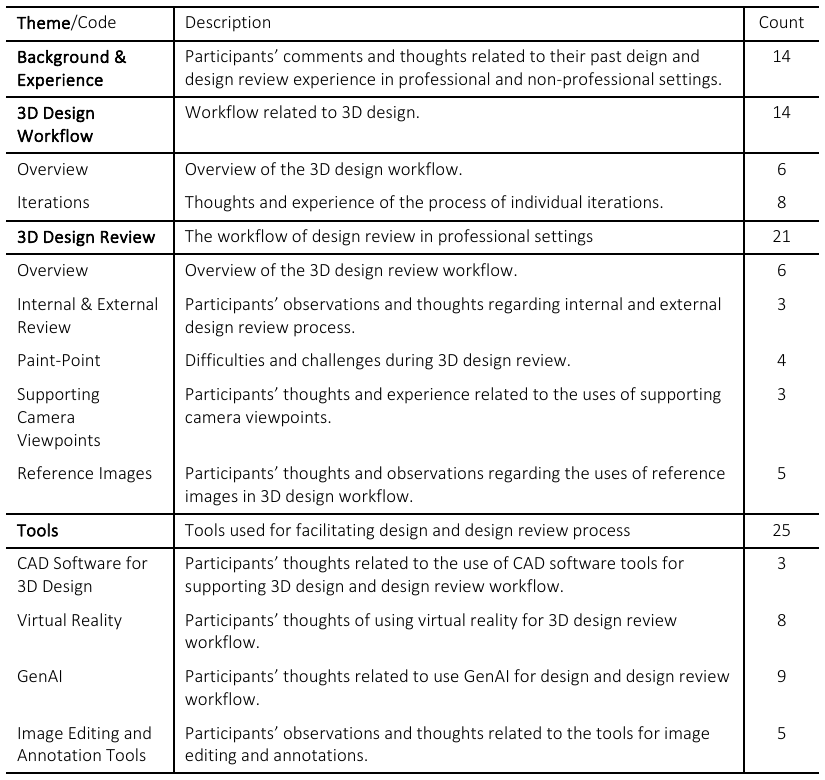}
    \caption{Themes and codes for Formative Study 1. Notably, it is possible that multiple codes are assigned to one quote.}
    \label{fig::codebook-online-dataa-nalysis-formative-study-1}
\end{figure*}

\begin{figure*}[t]
    \centering
    \includegraphics[width=0.65\textwidth]{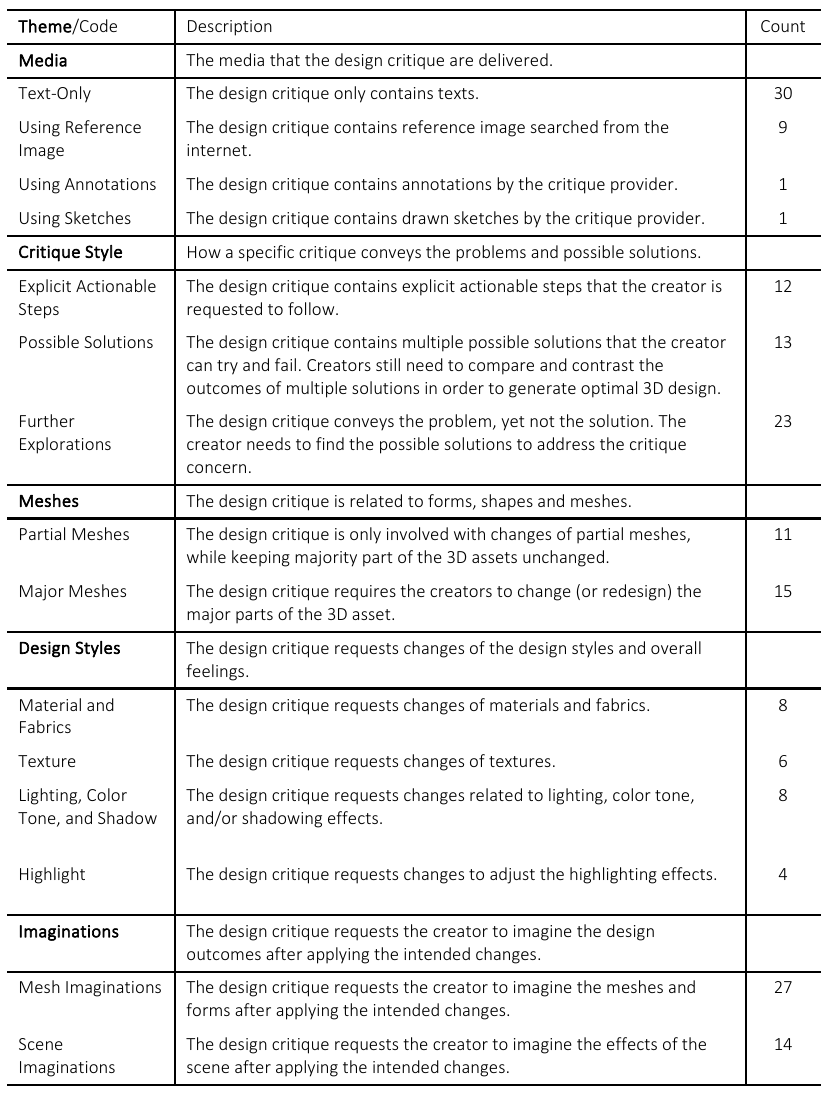}
    \caption{Themes and codes for analyzing real-world 3D design feedback data from Formative Study 2 . Notably, same feedback might be labelled by multiple codes.}
    \label{fig::codebook-online-data-analysis}
\end{figure*}

\begin{figure*}[t]
    \centering
    \includegraphics[width=\textwidth]{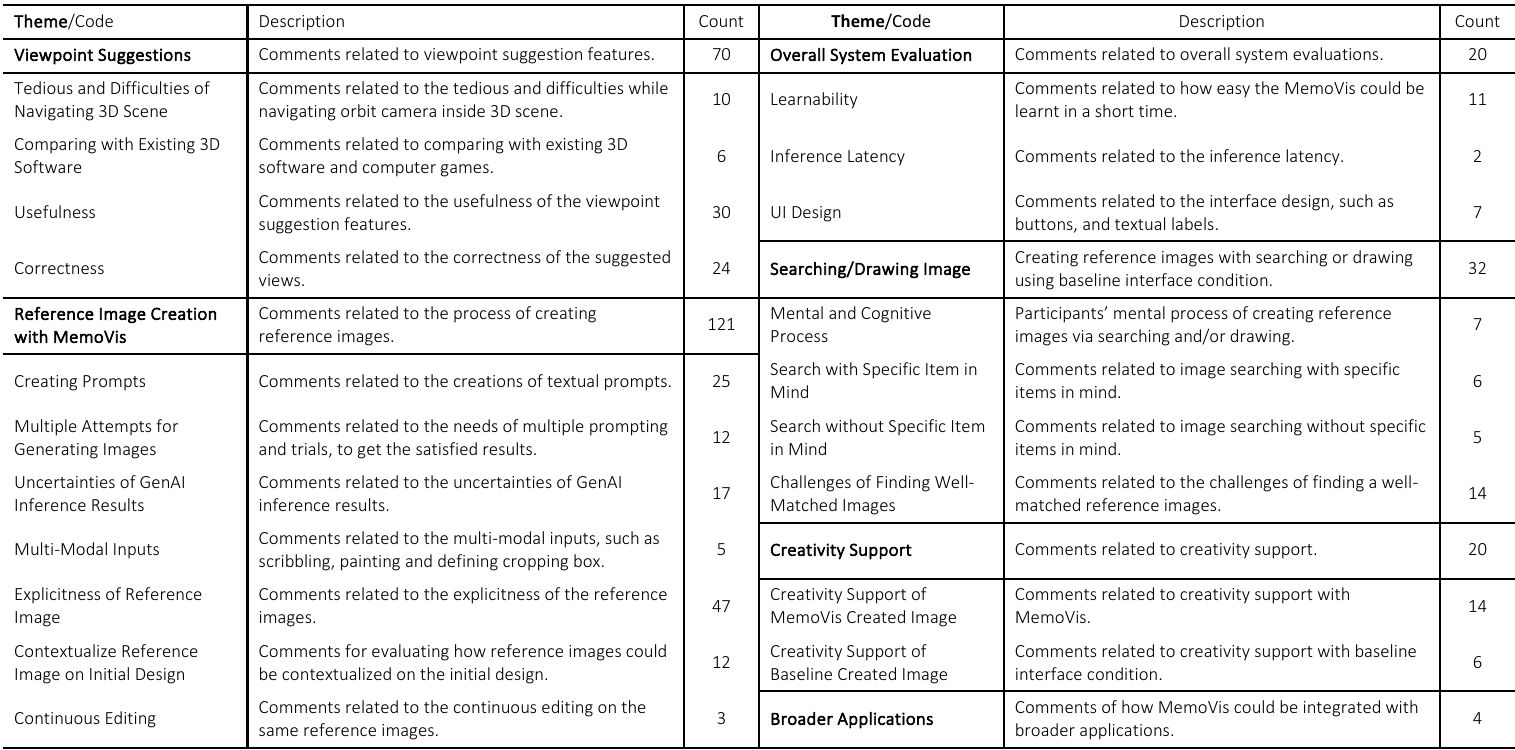}
    \caption{Themes and codes for Study 1. Notably, it is possible that multiple codes are assigned to one quote.}
    \label{fig::codebook-study-1}
\end{figure*}

\begin{figure*}[t]
    \centering
    \includegraphics[width=0.65\textwidth]{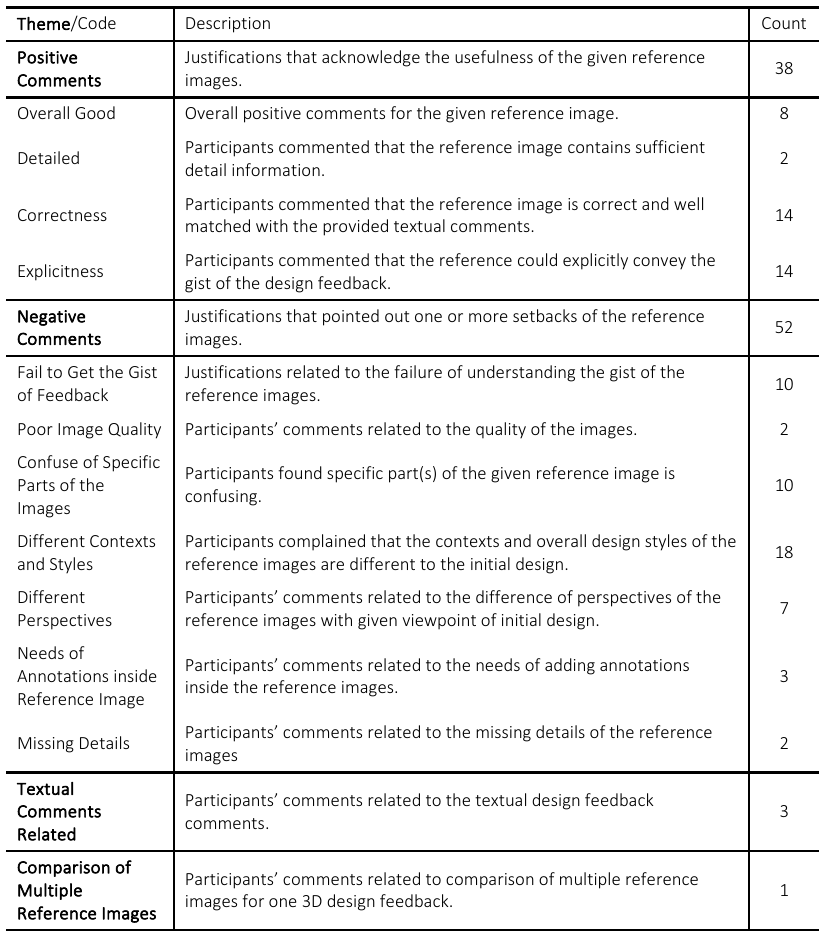}
    \caption{Themes and codes for participants' survey responses from Study 2. Notably, it is possible that multiple codes are assigned to one response.}
    \label{fig::codebook-study-2}
\end{figure*}

As part of supplementary material, we attached the resultant codebook for qualitative analysis for Formative Study 1 (Fig.~\ref{fig::codebook-online-dataa-nalysis-formative-study-1}), Formative Study 2 (Fig.~\ref{fig::codebook-online-data-analysis}), and final user studies with feedback provider participants (Fig.~\ref{fig::codebook-study-1}) and designer participants (Fig.~\ref{fig::codebook-study-2}).
Notably, multiple codes might be assigned to each observation (\eg~participants' quote, 3D design feedback, and survey responses from Study 2).

